\documentclass[12pt,letterpaper]{article}
\usepackage[affil-it]{authblk}
\usepackage[french,english]{babel}
\usepackage[T1]{fontenc}
\usepackage{authblk}
\usepackage[top=1.0in, bottom=1.0in, left=1in, right=1in]{geometry}
\usepackage{setspace}
\usepackage[utf8]{inputenc}
\usepackage{amsmath,tikz}
\usepackage{xcolor}
\usepackage{textcomp}
\usetikzlibrary{matrix}
\usepackage{amsfonts}
\usepackage{amssymb}
\usepackage{bbm}
\usepackage{graphicx}
\usepackage{subcaption}
\usepackage[labelformat=parens,labelsep=quad,skip=4pt,
					   format=plain, labelfont=it, textfont=it]{caption}
\usepackage{changepage}
\usepackage{longtable}
\usepackage{rotating}
\usepackage{array}
\usepackage{multicol}
\usepackage{multirow}
\usepackage{makecell}
\usepackage{pifont}
\usepackage{setspace}
\usepackage{hyperref}
\usepackage{natbib}
\usepackage{algorithm}
\usepackage{algpseudocode}
\usepackage{enumitem}

\usepackage{courier}
\usepackage{amsthm}
\usepackage{color}
\usepackage{comment}
\usepackage{latexsym}
\usepackage{verbatim}
\usepackage{bookmark}
\usepackage{pdfpages}
\usepackage{fix-cm}
\usepackage{etoolbox}

\definecolor{dodgerblue}{rgb}{0.12, 0.56, 1.0}
\definecolor{caribbeangreen}{rgb}{0.0, 0.8, 0.6}
\definecolor{blush}{rgb}{0.87, 0.36, 0.51}

\DeclareMathOperator*{\argmin}{arg\,min}

\hypersetup{
	colorlinks,
	linkcolor={dodgerblue},
	citecolor={caribbeangreen},
	urlcolor={blush}
}

\newtheorem{Rem}{Remark}

\newtheorem{The}{Theorem}

\newenvironment{Proof} {\noindent {\textbf{Proof}}} { \hfill $\Box$ \\ }
\newtheorem{Prop}{Proposition}
\newtheorem{Def}{Definition}

\title{\vspace{-75pt}Joint clustering with alignment for temporal data \\ \vspace{-20pt} in a one-point-per-experiment setting}

\author{Polina \textsc{Arsenteva}$^{1,2}$, Mohamed Amine \textsc{Benadjaoud}$^{3}$, Herv\'{e} \textsc{Cardot}$^{1}$ \\\vspace{-7pt}
	\footnotesize{1) Institut de Math\'{e}matiques de Bourgogne, UMR CNRS 5584, Universit\'{e} de Bourgogne, Dijon, France} \\ \vspace{-7pt}
		\footnotesize{2) IRSN PSE-SANTE/SERAMED/LRMed, Fontenay aux roses, France} \\\vspace{-7pt}
		\footnotesize{3) IRSN PSE-SANTE/SERAMED, Fontenay aux roses, France}}

\date{\vspace{-50pt}}

\begin{document}

\maketitle 
\begin{abstract}
Temporal data, obtained in the setting where it is only possible to observe one time point per experiment, is widely used in different research fields, yet remains insufficiently addressed from the statistical point of view. Such data often contain observations of a large number of entities, in which case it is of interest to identify a small number of representative behavior types.  In this paper, we propose a new method that simultaneously performs clustering and alignment of temporal objects inferred from these data, providing insight into the relationships between entities. Simulations confirm the ability of the proposed approach to leverage multiple properties of the complex data we target such as accessible uncertainties, correlations and a small number of time points. We illustrate it on real data encoding cellular response to a radiation treatment with high energy, supported with the results of an enrichment analysis. 
\end{abstract}

\section{Introduction}\label{sec:vitro_intro}
Temporal data are omnipresent in many scientific disciplines, including biology, physics, econometrics, and others. Among these cases, we encounter instances where longitudinal data are observed on the same individuals. In such cases, statistical inference can be addressed by frameworks such as functional data analysis \citep{ramsay_functional_2005}, time series and Gaussian process inference \citep{rasmussen_gaussian_2005}. However, it is often not feasible to perform measurements at different time points on the same individuals. For example, such context is common in biology, including in vitro studies, which often require destructive measurements such as real-time qPCR or microarray (e.g. \cite{gomez-cabrero_stategra_2019, luo_aloe-emodin_2023, sifakis_elucidating_2011}), and in vivo, which often necessitate the sacrifice of the animal (e.g. \cite{bertho_preclinical_2020, hu_inferring_2023, kuballa_moult_2011}). 

\begin{figure}
	\centering
	\begin{subfigure}{.5\textwidth}
		\centering
		\includegraphics[clip, trim=0cm 7.5cm 0cm 9cm, width=\textwidth]{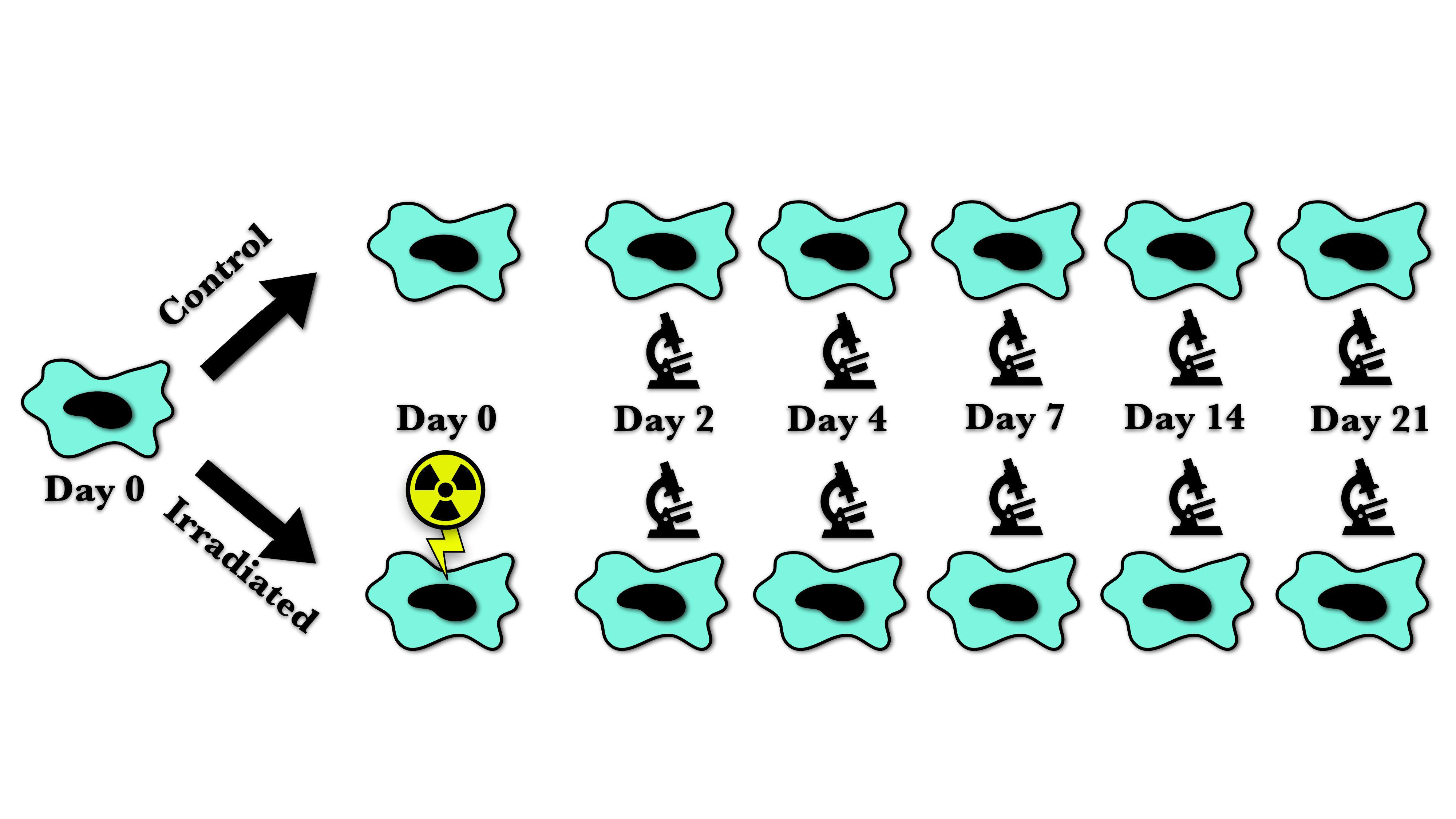}
		\caption{\label{fig:exp_design}}
	\end{subfigure}%
	\begin{subfigure}{.5\textwidth}
		\centering
		\includegraphics[clip,width=.75\linewidth]{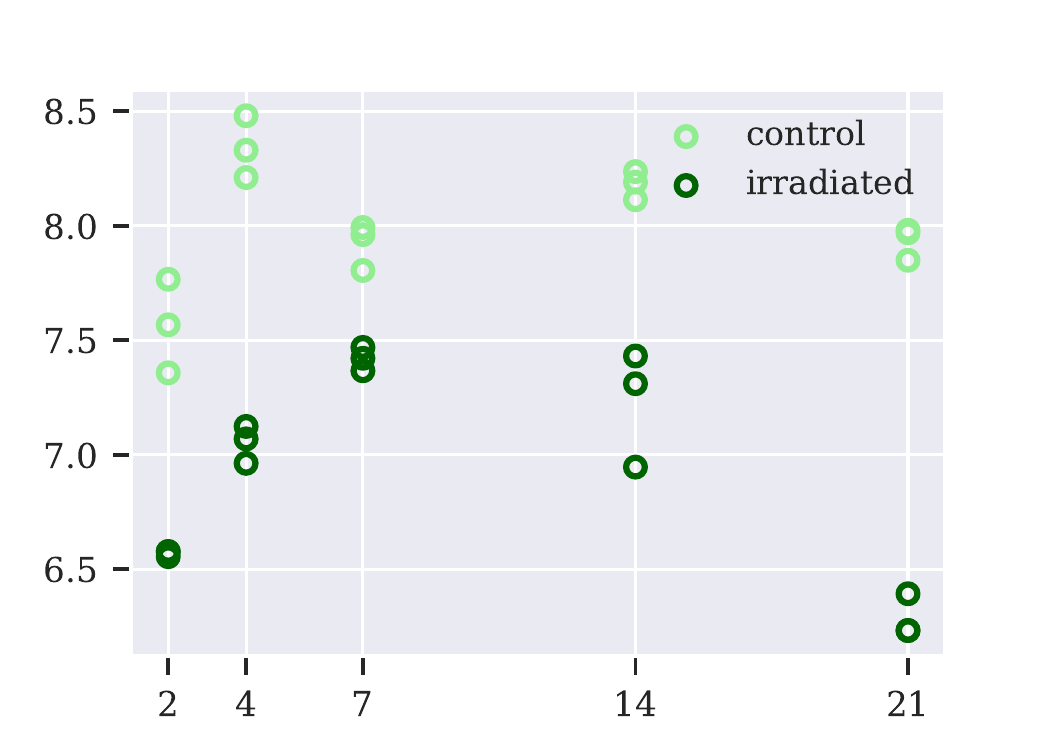}
		\caption{\label{fig:data_example}}
	\end{subfigure}
	\caption{\label{fig:data}A schematic representation of the experimental design used to obtain the data motivating this paper (a), along with a data plot with the expression of the gene VWF as an example (b). Transcriptional profiles of human endothelial cells were measured with real-time qPCR under control and under a single irradiation dose of 20 Gy at 0 h using a linear particle accelerator (LINAC) at 4 MV, 2.5 Gy/min, at 2, 4, 7, 14 and 21 days, with 3 replicates. Every time point and replicate is observed on a separate culture flask. For each culture flask, the expressions of 172 genes are measured simultaneously.}
\end{figure}

This paper is motivated by an in vitro experiment investigating the response of human endothelial cells to irradiation with a high energy level. Figure \ref{fig:data} schematically represents the experimental setting. Given the objective of studying the dynamics of the response to the treatment, multiple time points are measured. Due to the destructive nature of the measurements, which compromises the cells and renders them unsuitable for repeated measurements, each time point is observed on a separate culture flask. This precludes the possibility of observing temporal correlations in the dataset and treating the data as a time series. In practice, just prior to the experiments a single cellular population is separated into sub-populations, which are used to take measurements for each time point. Consequently, the flasks utilized for different time points contain different cells, which allows us to reasonably assume independence with respect to the temporal dimension.

The datasets considered in this paper share several other characteristics contributing to their complexity, leading to a non-trivial statistical problem. One important characteristic is the presence of two conditions, in  which case the focus is on studying the differences in response between the two conditions. In experiments, both treated and non-treated conditions (i.e. case and control) are necessary to study the effect of the treatment of interest. In the example presented in Figure \ref{fig:data}, these treatment conditions are irradiated and non-irradiated. The data from both conditions can be expressed as a fold change, representing the difference between the corresponding responses. Fold changes are commonly used in biology, but are generic by definition and can be applied in different fields. For example, in econometrics, a similar concept is referred to as a "treatment effect" \citep{wooldridge_econometric_2010}.

Furthermore, multiple replicates are necessary to account for measurement uncertainties, which is especially crucial given that different time points are observed independently. In this case, the mean values across replicates represent an actual signal, but they are insufficient for inference on their own, as they do not include information on uncertainties. Consequently, the inference approach must consider the available information on uncertainties and cross-sectional correlations between entities. This implies the inadequacy of simple techniques such as only treating average signals, commonly seen in literature when performing clustering on such data (e.g. \cite{dang_dynamic_2017, kuballa_moult_2011, ren_comparative_2022, sifakis_elucidating_2011, shahid_comparative_2019}). It is common practice to take uncertainties into account solely for the purpose of visualizing a limited number of entities (e.g. \cite{dang_dynamic_2017, luo_aloe-emodin_2023, shahid_comparative_2019, ren_comparative_2022}).

Finally, the aim of these studies is to assess the response of multiple entities, often outnumbering the time points that are identical across all entities. This makes functional data analysis and stochastic process-based approaches unsuitable for inference. The analysis in such cases typically involves reducing the datasets to a small number of groups, characterized by typical behavior templates. This methodologically translates into the task of clustering. Additional information about the joint distribution of entities is often available in the data and must be accounted for. To illustrate, the data presented in this study is transcriptomic, involving the measurement of the expression of hundreds of genes. For a given experimental condition and time point, the measurements for all genes are taken simultaneously from the cells of one culture flask, providing access to correlations between the genes. Additionally, in many cases, there exists prior knowledge of underlying causal relationships between the considered entities. In this regard, the temporal aspect of the data can be utilized by integrating the alignment into the clustering process. This approach has two key benefits: it enhances the clustering itself and provides insights into the predictive nature of the entities. There are several approaches to joint clustering with alignment in the context of functional data analysis (e.g. \cite{sangalli_k-mean_2010, cremona_probabilistic_2023}) and Gaussian process inference (e.g. \cite{kazlauskaite_gaussian_2019}). While these are not directly applicable to the data targeted in this paper, they provide useful inspiration for the proposed approach.

In summary, our objective is to construct a joint clustering with alignment procedure for complex data that exhibits the following challenging features: a small number of time points that are observed on different individuals, two experimental conditions, multiple replicates providing information on cross-section correlations and uncertainties, and a large number of entities to cluster. In this paper, we present a statistical framework that addresses these challenges while exploiting data properties to ensure efficient computation. 

In Section \ref{sec:clust}, we introduce fold change estimators adapted to the aforementioned properties of time points and the presence of replicates. Focusing on fold changes allows to account for the two experimental conditions present. We introduce a procedure that performs simultaneous alignment and clustering of the fold changes based on dissimilarities constructed as a generalization of the $L^2$-distance between Gaussian variables. Combined with these dissimilarities, the chosen clustering procedure allows to fully leverage the information available through replicates. The alignment procedure integrated in the clustering is designed for a small number of time points that are measured independently. We show that the proposed algorithm has a reduced complexity compared to its state-of-the-art counterpart, which is advantageous when dealing with a high number of entities. The proposed approach is evaluated in the framework of multiple simulation studies in Section \ref{sec:sim} in comparison with a number of alternative choices. Finally, in Section \ref{sec:applic} we present an application to the data acquired in the radiobiological experimental setting presented earlier. The obtained information is validated through the bioinformatic enrichment analysis of the clusters.

\section{Materials and methods}\label{sec:clust}
\subsection{Fold change modeling and estimation}\label{sec:estim}
To avoid introducing non-existent information by smoothing the temporal response, we consider time as discrete. For a given dataset, we define the response variable as $Y_{i,k,j,t}$ for an entity (gene) $i \in \{1,2,\dotso,n_e\}$, under the experimental condition $k=0$ if control (non-irradiated) and $k=1$ if case (irradiated), at a time point $t \in \{t_1,t_2,\dotso,t_p\}$ and for a replicate $j \in \{1,2,\dotso,n_r\}$. We assume there are $n_r$ observations for both experimental condition and every time point without loss of generality, given that $n_r \geq 2$. 

Two constraints with respect to the covariances between the responses follow from the specificity of the experimental design. On the one hand, the measures of expressions for all genes for a given experimental condition and time point are collected from the same plate, which allows to estimate cross section covariances between genes, i.e. $\mathrm{Cov}\left(Y_{i,k,j,t}, Y_{i',k,j,t} \right)$ for $i \neq i'$. On the other hand, we do not have access to the temporal covariance structure due to the destructive technique used in collecting measures from a plate for a given time point. Thus, measures for different time points are produced individually on different cells and are not correlated, i.e. given distinct time points $t \neq t'$, for any replicate pair $(j, j') \in \{1,2,\dotso,n_r\}^2$ and entity pair $(i, i') \in \{1,2,\dotso,n_e\}^2$ we have $\mathrm{Cov}\left(Y_{i,k,j,t}, Y_{i',k,j',t'} \right) = 0$.

Classical estimators of the fold changes in the multivariate setting are the pointwise estimators: a set of empirical individual fold changes is denoted by $\Gamma = \left(\Gamma_1, \dotso, \Gamma_{n_e}\right)$ where $\Gamma_i = \left(\Gamma_{i,t_1}, \dotso, \Gamma_{i,t_p}\right)$ such that $\Gamma_{i,t} = \frac{\sum_{j=1}^{n_r}Y_{i,1,j,t} - \sum_{j=1}^{n_r}Y_{i,0,j,t}}{n_r} = \overline{Y_{i,1,t}} - \overline{Y_{i,0,t}}$, representing the difference between the means of the control and the case response. However, these estimators do not take into account the information about uncertainties and correlations present in the data. We propose a new definition of fold change estimators to fully account for all information about their estimated distributions:

\begin{Def}\label{def:fc}
	The estimator of the fold change of entity $i$ is denoted by $\widehat{\Gamma}_i$, assumed to be a random Gaussian vector and is defined as $\widehat{\Gamma}_i  | \Gamma_i, \Sigma_{\Gamma_i} \sim \mathcal{N}\left(\Gamma_i, \Sigma_{\Gamma_i}\right)$ where  $\Gamma_{i,t} = \overline{Y_{i,1,t}} - \overline{Y_{i,0,t}}$ (the pointwise estimator), and $\Sigma_{\Gamma_i}$ is a diagonal matrix with the diagonal $(\sigma^2_{\Gamma_{i,t_1}}, \dots, \sigma^2_{\Gamma_{i,t_p}})$ such that $\sigma^2_{\Gamma_{i,t}} = \sigma^2_{Y_{i,1,t}}  + \sigma^2_{Y_{i,0,t}}$ is the sum of unbiased sample variances of $(Y_{i,0,j,t})_{1 \leq j \leq n_r}$ and $(Y_{i,1,j,t})_{1 \leq j \leq n_r}$, with $\sigma^2_{Y_{i,k,t}} = \frac{\sum_{j=1}^{n_r} (Y_{i,k,j,t} - \overline{Y_{i,k,t}} )^2}{n_r - 1}$ for $k \in \{0,1\}$.
\end{Def} 

\begin{Rem}
	The fact that the covariance matrix in Definition \ref{def:fc} is diagonal is a direct consequence of the second covariance constraint mentioned above.
\end{Rem}

\subsection{Distance between fold change estimators}\label{sec:dist}
Since the task at hand is clustering of the estimators of fold changes, and thus distribution clustering, there is a need to choose an appropriate distance. First, we expand the Definition \ref{def:fc} to a pair of fold changes by specifying their joint distribution: 
\begin{Def}\label{def:fc_pair}
	The estimator of a pair of fold changes of entities $i$ and $i'$ is denoted as $\left[ \widehat{\Gamma}_i ^\intercal  \widehat{\Gamma}_{i'} ^\intercal \right]^\intercal $, assumed to be a random Gaussian vector and is defined as follows: 
	$$
	\begingroup 
	\renewcommand*{\arraystretch}{0.7}
		\begin{bmatrix} \widehat{\Gamma}_i \\  \widehat{\Gamma}_{i'} \end{bmatrix}\sim \mathcal{N}\left(\begin{bmatrix} \Gamma_i \\ \Gamma_{i'} \end{bmatrix}, \begin{bmatrix} 
			\Sigma_{\Gamma_i} & \mathrm{P}_{\Gamma_i \Gamma_{i'}} \\
			\mathrm{P}_{\Gamma_i \Gamma_{i'}}^\intercal & \Sigma_{\Gamma_{i'}}
		\end{bmatrix} \right),
	\endgroup
	$$
	where the quantities $\Gamma_i$, $\Gamma_{i'}$, $\Sigma_{\Gamma_i}$ and $\Sigma_{\Gamma_{i'}}$ describing marginal distributions of $\widehat{\Gamma}_{i}$ and $\widehat{\Gamma}_{i'}$ are defined according to Definition \ref{def:fc}, and the diagonal cross-covariance matrix $\mathrm{P}_{\Gamma_i \Gamma_{i'}}$ with the diagonal $(\rho_{\Gamma_{i,t_1} \Gamma_{i',t_1}}, \dots, \rho_{\Gamma_{i,t_p} \Gamma_{i',t_p}})$ such that $\rho_{\Gamma_{i,t} \Gamma_{i',t}} = \rho_{Y_{i,1,t} Y_{i',1,t}} + \rho_{Y_{i,0,t} Y_{i',0,t}}$ is the sum of sample covariances of $\left((Y_{i,0,j,t},Y_{i',0,j,t})\right)_{1 \leq j \leq n_r}$ and $\left((Y_{i,1,j,t},Y_{i',1,j,t})\right)_{1 \leq j \leq n_r}$, with $\rho_{Y_{i,k,t} Y_{i',k,t}} = \frac{\sum_{j=1}^{n_r} (Y_{i,k,j,t}  - \overline{Y_{i,k,t} } )(Y_{i',k,j,t} - \overline{Y_{i',k,t}})}{n_r - 1}$ for $k \in \{0,1\}$.
\end{Def} 

We use the $L^2$-distance between Gaussian random variables, since it is the simplest distance that allows to account for the information on the joint distributions of the random variables. In comparison, most other well-known distances, such as total variation and Hellinger, only take the marginal distributions into account. The squared $L^2$-distance between fold change estimators $\widehat{\Gamma}_{i}$ and  $\widehat{\Gamma}_{i'}$, with the joint distribution given in Definition \ref{def:fc_pair}, will be denoted as $\mathbf{\widehat{d_2^2}}$ and can be calculated using a well-known formula as follows: 
\begin{equation}
\begin{split}
	\mathbf{\widehat{d_2^2}}\left(\widehat{\Gamma}_i,\widehat{\Gamma}_{i'} \right) &= \mathbb{E}\Vert \widehat{\Gamma}_i- \widehat{\Gamma}_{i'} \Vert ^2_2 = \Vert \Gamma_i -\Gamma_{i'} \Vert ^2_2 + \mathrm{Tr} (\Sigma_{\Gamma_i}) + \mathrm{Tr} (\Sigma_{\Gamma_{i'}}) - 2 \mathrm{Tr} ( \mathrm{P}_{\Gamma_i \Gamma_{i'}}) \\
	&= \sum_{l=1}^{p} \left( \Gamma_{i,t_l} - \Gamma_{i',t_l} \right)^2 + \sum_{l=1}^{p} \sigma^2_{\Gamma_{i,t_l}} + \sum_{l=1}^{p} \sigma^2_{\Gamma_{i',t_l}} - 2 \sum_{l=1}^{p} \rho_{\Gamma_{i,t_l} \Gamma_{i',t_l}},
\end{split}
\end{equation}
where $\Vert \cdot \Vert_2$ is the Euclidean norm. 

\subsection{Fold change alignment}\label{sec:align}
In this section, we introduce all the mathematical quantities necessary to perform the temporal alignment of the fold changes, which will be further applied jointly with clustering. First, we define a transformation of a pair of time vectors that will be referred to as a time warp. In contrast with a similar concept in functional data analysis, we define time warp as a shift of time vector indices, rather than time vectors themselves.

\begin{Def}\label{def:tw}
	Let $\mathbf{t}  = (t_1, \dots, t_p)  \in \mathbbm{R}$ be a time vector for the considered dataset. A time warp $\mathcal{W}_s$ of step $s \in \mathbbm{Z}$ is a transformation of two time vectors, defined as follows:
	\begin{align*}
		\mathcal{W}_s \colon  \mathbbm{R}^{p} &\to  \mathbbm{R}^{2(p-|s|)}\\
		 \mathbf{t} & \mapsto \mathbf{t^{	\mathcal{W}_s}} =  (\mathbf{t_{1}},  \mathbf{t_{2}} )
		\text{ where: } 
	\end{align*} 
	$$
	\mathbf{t_{1}} = \begin{cases}\begin{aligned}
		(t_1, \dots, t_{p-|s|}) \text{ if } s>0 \\[-10pt]
		(t_{1+|s|}, \dots, t_{p})\text{ if } s<0 \\[-10pt]
		 \mathbf{t}  \text{ if } s=0
		 \end{aligned}
	\end{cases}\
	\text{ and } \hspace{15pt}
	\mathbf{t_{2}} = \begin{cases}\begin{aligned}
		(t_{1+|s|}, \dots, t_{p}) \text{ if } s>0 \\[-10pt]
		(t_1, \dots, t_{p-|s|})  \text{ if } s<0 \\[-10pt]
		 \mathbf{t}  \text{ if } s=0
		 \end{aligned}
	\end{cases}.\
	$$
\end{Def}

\begin{figure}
	\centering
	\includegraphics[clip, trim=1cm 0.5cm 1cm 1.5cm,width=0.6\textwidth]{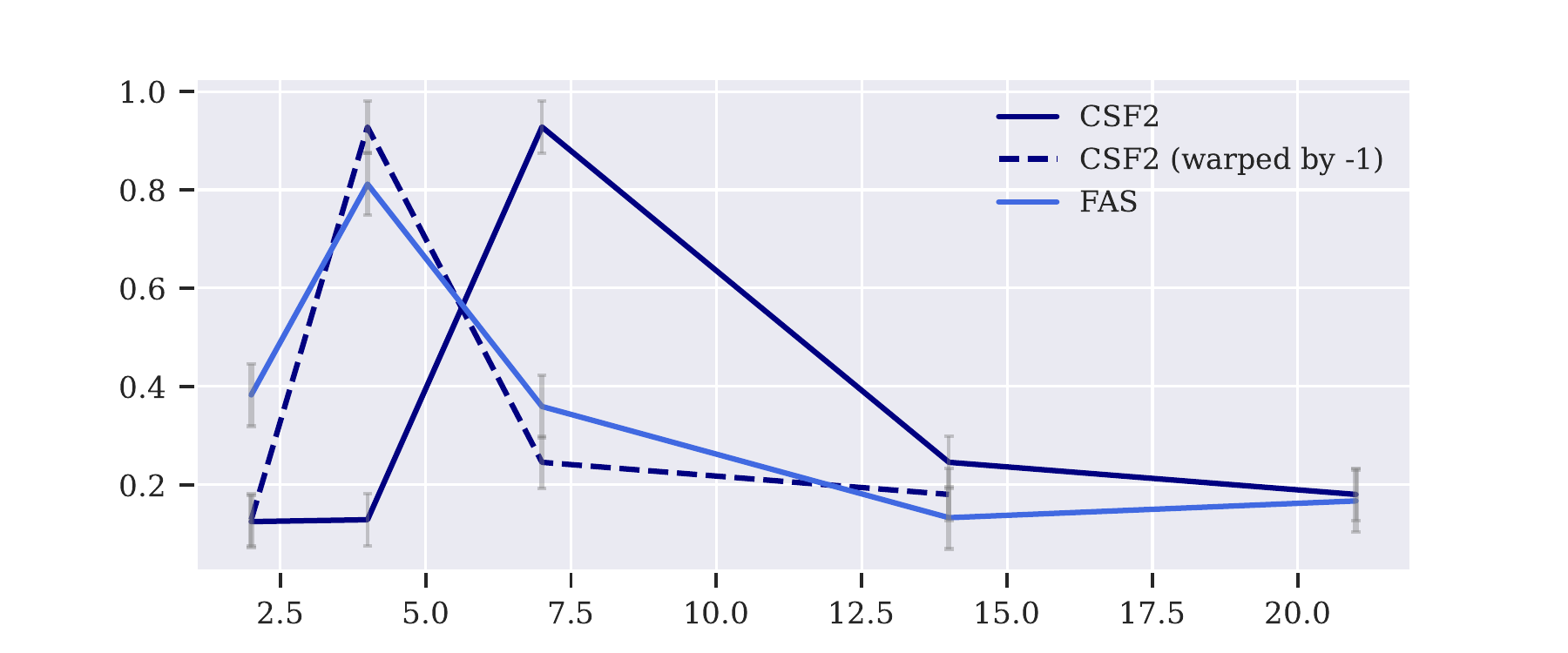}
	\caption{\label{fig:tw_example}The effect of time warping illustrated on a figure, where means with standard deviation of a pair of normalized transcriptomic fold changes are plotted. It can be observed that after warping the fold change of the gene CSF2 backwards, its mean practically coincides with that of the fold change of the gene FAS. In this example, the quantities given in Definition \ref{def:tw} are as follows: $\mathbf{t}=(2, 4, 7, 14, 21)$, $\mathbf{t_{1}}=(4, 7, 14, 21)$ and $\mathbf{t_{2}}=(2, 4, 7, 14)$.}
\end{figure}

In this definition, we distinguish three major warping types: backward warp ($s<0$), forward warp ($s>0$) and identity warp ($s=0$). The concept of alignment is illustrated in Figure \ref{fig:tw_example}. In this example, the fold changes are highly comparable up to a time shift, indicating that alignment should markedly reduce the distance between them and thus categorize them in the same cluster. It can be noted that our alignment framework has similarities with Dynamic Time Warping (DTW). For two temporal sequences, the alignments obtained with our approach can be represented on the DTW "distance" matrix in the form of paths on the sub- and superdiagonals of the matrix\footnote{The constraint that the first index from the first sequence must match the first index from the other sequence, and similarly for the last index pair, must be relaxed.}. However, our approach differs from DTW in that we consider simultaneous shifts of all time points in the same direction with the same step. Then dissimilarities between entire fold changes are calculated, rather than point-wise dissimilarities, which allows to account for the information obtained with multiple replicates.

Next, we define a warped fold changes pair in terms of the original fold changes:

\begin{Def}\label{def:warped_fc_pair}
	Let $ \mathbf{t} \in  \mathbbm{R}^{p}$ be a $p$-dimensional time vector, and $s \in \mathbbm{Z}$ a warp step. We denote as $\left[ \widehat{\Gamma_{i} ^{\mathcal{W}_s} } ^\intercal  \widehat{\Gamma_{i'} ^{\mathcal{W}_s} } ^\intercal \right]^\intercal$ an $s$-warped fold changes pair $\left[ \widehat{\Gamma}_i ^\intercal  \widehat{\Gamma}_{i'} ^\intercal \right]^\intercal $ such that:
	$$
	\begin{bmatrix} \widehat{\Gamma_i ^{\mathcal{W}_s} } \\ \widehat{\Gamma_{i'} ^{\mathcal{W}_s} }  \end{bmatrix} =
	 \begin{bmatrix}\left( \widehat{\Gamma}_i^{t^{\mathcal{W}_s}_1}, \dotso,  \widehat{\Gamma}_i^{t_{1,p-|s|}^{\mathcal{W}_s}} \right)^\intercal  \\ \left( \widehat{\Gamma}_{i'} ^{t_{2, 1}^{\mathcal{W}_s}}, \dotso,  \widehat{\Gamma}_{i'} ^{t_{2,p-|s|}^{\mathcal{W}_s}} \right)^\intercal  \end{bmatrix},
	$$
	where  $\mathbf{t_1^{\mathcal{W}_s}} = (t_{1, 1}^{\mathcal{W}_s} \dots, t_{1, p-|s|}^{\mathcal{W}_s})$ and $\mathbf{t_1^{\mathcal{W}_s}} = (t_{2, 1}^{\mathcal{W}_s} \dots, t_{2, p-|s|}^{\mathcal{W}_s})$ such that $(\mathbf{t_1^{\mathcal{W}_s}} , \mathbf{t_2^{\mathcal{W}_s}}) =\mathcal{W}_s( \mathbf{t} )$.
\end{Def}

For every fold changes pair only the first fold change is being moved since it allows for a more convenient manipulation of warping results while being able to examine all warping possibilities if considering both forward and backward type warping. According to the definitions presented above, the first fold change in the pair is being warped with a subsequent cutoff of extraneous parts. The second fold change in the pair does not move, however its parts that do not correspond to remaining post-warping points of the first one are also being cut off. The calculations are detailed in the proof of Proposition \ref{prop:diss}.

We introduce a new dissimilarity measure between the random fold changes estimators that is a generalization of the distance $\mathbf{\widehat{d^2_2}}$ in order to take all the covariances into account in the case where time warping is applied:

\begin{Def}\label{def:diss}
	Let $s \in \mathbbm{Z}$, and $X$ and $Y$ be $p$-dimensional Gaussian random variables with a joint distribution 
	$$
	\begingroup 
	\renewcommand*{\arraystretch}{0.7}
	\begin{bmatrix} X\\Y  \end{bmatrix} \sim \mathcal{N}\left(\begin{bmatrix}\mu_X \\ \mu_Y  \end{bmatrix}, \begin{bmatrix} 
		\Sigma_X & \mathrm{P}_{XY} \\
		\mathrm{P}_{XY}^\intercal & \Sigma_{Y}
	\end{bmatrix} \right) \text{ such that }
	\left[\mathrm{P}_{XY} \right]_{ij} = 0  \text{ if } j \neq i-s.
	\endgroup
	$$
	We define a dissimilarity measure $\mathbf{\widehat{diss}_s}$ between $X$ and $Y$ as follows:
	$$
		\mathbf{\widehat{diss}_s} \left(X, Y \right) 
		=  \Vert \mu_X -\mu_Y \Vert ^2 + \mathrm{Tr} (\Sigma_X) + \mathrm{Tr} (\Sigma_Y) 
		- 2 \sum_{l=1}^{p-|s|} \left[\mathrm{P}_{XY} \right]_{\left(l+s \mathbbm{1}_{\mathbbm{Z}_{+}^{*}}(s),  l+s \mathbbm{1}_{\mathbbm{Z}_{-}^{*}}(s)\right)} .
	$$
\end{Def}

Applying the dissimilarity measure to the warped fold changes, we get the expression resembling the value of  $\mathbf{\widehat{d^2_2}}$ in the case of non-warped fold changes, with extraneous part getting cut off as a result of warping:

\begin{Prop}\label{prop:diss}
	Let $s \in \mathbbm{Z}$ be a warp step, and $\left[ \widehat{\Gamma_{i} ^{\mathcal{W}_s} } ^\intercal  \widehat{\Gamma_{i'} ^{\mathcal{W}_s} } ^\intercal \right]^\intercal$ an $s$-warped fold changes pair. The value of dissimilarity $\mathbf{\widehat{diss}_s}$ between the fold changes $\widehat{\Gamma_i ^{\mathcal{W}_s} }$ and $\widehat{\Gamma_{i'}^{\mathcal{W}_s} }$ can be expressed in the following form:
$$
	\mathbf{\widehat{diss}_s} \left(\widehat{\Gamma_i ^{\mathcal{W}_s} }, \widehat{\Gamma_{i'} ^{\mathcal{W}_s} }\right) 
	= \sum_{l=l^{*}}^{p^{*}} \left( \Gamma_{i,t_l} - \Gamma_{i',t_{l+s}} \right)^2 + \sum_{l=l^{*}}^{p^{*}} \sigma^2_{\Gamma_{i,t_l}} + \sum_{l=l^{*}}^{p^{*}} \sigma^2_{\Gamma_{i',t_{l+s}}} - 2 \sum_{l=1+|s| }^{p-|s|} \rho_{\Gamma_{i,t_l} \Gamma_{i',t_l}},
$$
where $l^{*}=1-s \mathbbm{1}_{\mathbbm{Z}_{-}^{*}}(s)$ and $p^{*}=p-s \mathbbm{1}_{\mathbbm{Z}_{+}^{*}}(s)$.
\end{Prop}

\begin{Rem}
	 It can be noted that $\mathbf{\widehat{diss}_s}\left(\widehat{\Gamma_i ^{\mathcal{W}_s} }, \widehat{\Gamma_{i'} ^{\mathcal{W}_s} }\right)=\mathbf{\widehat{d_2^2}}\left(\widehat{\Gamma}_i,\widehat{\Gamma}_{i'} \right)$ in the case of the identity warp $s=0$.
\end{Rem}

\begin{Rem}\label{rem:cutoff}
	We opted to construct time warping in a way that some parts of the fold changes that move outside of the temporal domain under consideration are excluded. This approach was selected over alternatives based on extending the fold changes instead of cutting them, because it avoids the introduction of unobserved information. However, this approach may introduce a bias in comparisons between warped and unwarped sequences. To ensure a fair and accurate assessment, it is essential to normalize the dissimilarities with respect to the number of post-warping time points in order to render them comparable (implemented in our package \textit{scanfc}).
\end{Rem}

\subsection{Joint clustering with alignment}\label{sec:cl_n_al}
Our objective is to reduce the fold changes to a small number of behavior types up to a time shift, which translates into clustering of aligned fold changes. We perform clustering using a "k-means like" version of k-medoids \citep{park_simple_2009} based on a series of random initializations of type k-means++ \citep{arthur_k-means_2007}. The k-medoids algorithm was selected due to a number of properties that are beneficial in the context discussed in this paper. These properties stem from the fact that clustering is performed by comparing elements to a medoid, which is itself an element of the dataset. In addition to enhanced robustness and interpretability, it also provides a straightforward and efficient approach to preserving the information on joint distributions of fold change pairs throughout the clustering process when combined with the dissimilarity from Definition \ref{prop:diss}. Furthermore, using k-medoids allows for alignment to be clustering-dependent: while elements are compared to medoids for clustering, their warps can also be chosen in a unique way with respect to medoids. Pseudocode for the state-of-the-art version of joint clustering and alignment, applied in the context of this work, is presented in Algorithm \ref{alg:cl_n_al_iter} in Supplementary Material. 

We propose a modification of the algorithm that reduces the computation time by leveraging the low temporal dimensionality of the data in the multivariate setting. Given the typically small number of time points, the number of possible warps is necessarily even smaller and known in advance. Furthermore, the distributions of the fold change pairs under different warps are known, allowing for the calculation of all alignment options before clustering. This reduces computation time and is a non-memory-intensive process. In light of this, we introduce the following quantities, which will be used in the modified version of the algorithm:

\begin{Def}\label{def:owd}
	Let $\mathcal{S} = \{-s_{max}, \dots, s_{max}\} \subset \mathbbm{Z}$ be a finite set of allowed warp steps, given a maximal warping step $s_{max} \in  \mathbbm{N}$. The Optimal Warping Dissimilarity matrix, denoted $\mathcal{OWD}$, is a matrix containing the values of the dissimilarity measure $\mathbf{\widehat{diss}_s}$ for all pairs of fold changes in case of their optimal pairwise alignment over the set of all possible warps with steps in $\mathcal{S}$, or formally:
	$$
	\mathcal{OWD}=\left[\min_{s \in \mathcal{S}} \left[ \mathbf{\widehat{diss}_s}\left(\widehat{\Gamma_i ^{\mathcal{W}_s} }, \widehat{\Gamma_{i'} ^{\mathcal{W}_s} }\right) \right] \right]_{1 \leq i, i' \leq n_e}.
	$$
\end{Def}

\begin{Def}\label{def:ow}
	Let $\mathcal{S} = \{-s_{max}, \dots, s_{max}\} \subset \mathbbm{Z}$ be a finite set of allowed warp steps, given a maximal warping step $s_{max} \in  \mathbbm{N}$. The Optimal Warp matrix, denoted $\mathcal{OW}$, is a matrix containing, for all pairs of fold changes, the values in $\mathcal{S}$ corresponding to the warp steps allowing to achieve their optimal pairwise alignment with respect to the dissimilarity measure $\mathbf{\widehat{diss}_s}$, or formally:
	$$
	\mathcal{OW}=\left[\argmin_{s \in \mathcal{S}} \left[  \mathbf{\widehat{diss}_s}\left(\widehat{\Gamma_i ^{\mathcal{W}_s} }, \widehat{\Gamma_{i'} ^{\mathcal{W}_s} }\right) \right] \right]_{1 \leq i, i' \leq n_e}.
	$$
\end{Def}

\begin{Prop}\label{prop:sym}
	The following statements are true for matrices $\mathcal{OWD}$ and $\mathcal{OW}$:
	\begin{enumerate}
		\item  $\mathcal{OWD}$ is symmetric.
		\item $\mathcal{OW}$ is anti-symmetric.
	\end{enumerate}
\end{Prop}

\begin{Rem}
	$\mathcal{OW}$ allows to interpret the main warping types.  For a given fold changes pair,  if the optimal warp is the identity warp,  they are referred to as simultaneous.  If not,  then one fold change in the pair is warped forward with respect to the other,  whereas the other fold change is being warped backwards with respect to the first.  In this case,  the fold change that is warped forward is referred to as 'predictive' of other one,  whereas the latter is labeled as 'predicted', or 'regulated'. 
\end{Rem}

\begin{algorithm}
	\linespread{0.9}
	\footnotesize
	\caption{Joint clustering and alignment algorithm based on $\mathcal{OWD}$ and $\mathcal{OW}$ matrices}\label{alg:cl_n_al}
	\begin{algorithmic}[1]
		\Require Fold changes  $\widehat{\Gamma}=\left(\widehat{\Gamma}_1, \dots, \widehat{\Gamma}_{n_e}\right)$, $K \in \mathbbm{N}$, $it_{max} \in \mathbbm{N}$, $n_{init} \in \mathbbm{N}$, $\epsilon>0$.
		\State Compute $\mathcal{OWD}$ and $\mathcal{OW}$
		\State $TC \gets \infty$ \Comment{global total cost} 
		\For{$init \in \{1, \dots, n_{init} \}$}
		\State Initialize centroids $C = \left(C_1 , \dots,  C_K \right) \subset  \{1 , \dots,  n_e\} $ with kmeans++
		\State $TC_{it} \gets \infty$ \Comment{total cost of the current initialization} 
		\State $\Delta TC \gets \infty$ \Comment{change in total cost of the current initialization} 
		\State $it \gets 1$
		\While{$\Delta TC > \epsilon$ \textbf{and} $it < it_{max}$}
		\State  $TC_{it}^{new} \gets 0$  
		
		\State \textbf{1. Assign step:}
		\For{$i \in\{1 , \dots,  n_e\}$}
		\State $Cl_i \gets \argmin_{k \in \{1 , \dotso,  K\} } \mathcal{OWD}_{iC_k}$ \Comment{Assign aligned fold changes to centroids} 	\label{lst:assign}
		\EndFor 
		
		\State  \textbf{2. Update step:}
		\For{$k \in \{1, \dots, K \}$} 
		\State $d_{min} \gets \infty $
		\For{$i \in cluster_k = \{ i \in \{1 , \dotso,  n_e\} | Cl_i=k\}$} \Comment{Candidate for a centroid}
		\State $d_{cluster_k}  \gets \sum_{i' \in cluster_k}\mathcal{OWD}_{ii'}$ \label{lst:update_first}
		\If{$d_{cluster_k} < d_{min}$} \Comment{Choose new centroid}
		\State $d_{min} \gets d_{cluster_k} $
		\State $C_k^{new}  \gets i $
		\EndIf \label{lst:update_last}
		\EndFor
		\State  $TC_{it}^{new} \gets TC_{it}^{new} + d_{min} $  
		\EndFor
		
		\State   \textbf{3. Calculate the change in total cost: } 
		\State  $\Delta TC \gets TC_{it} -  TC_{it}^{new}$ 
		\If{$\Delta TC  > \epsilon$}
		\State $C \gets \left(C_1^{new} , \dots,  C_K^{new} \right) $
		\State $TC_{it} \gets TC_{it}^{new}$
		\EndIf
		\State $it \gets it + 1$
		\EndWhile
		\If{$TC_{it} < TC$}
		\State $C \gets  \left(C_1 , \dots,  C_K \right)$\Comment{centroids labels}
		\State $Cl \gets  \left(Cl_1 , \dots,  Cl_{n_e} \right)$\Comment{cluster labels}
		\State $\mathcal{W} = \left(\mathcal{OW}_{1Cl_1} , \dotso,  \mathcal{OW}_{n_eCl_{n_e}} \right)$\Comment{warps} \label{lst:warps}
		\State $TC \gets TC_{it} $
		\EndIf
		\EndFor
		\State \Return $C$, $Cl$,  $\mathcal{W}$
	\end{algorithmic}
\end{algorithm}

The modified version of the previous algorithm, presented in Algorithm \ref{alg:cl_n_al}, is based on integrating time warping in the clustering process through pre-calculated matrices $\mathcal{OWD}$ and $\mathcal{OW}$. The algorithm is analogous to the original k-means like k-medoids (without the alignment), with the standard dissimilarity matrix being replaced with the matrix $\mathcal{OWD}$, containing information about the alignments. Following each initialization of the centroids, the algorithm iterates between "Assign" (assigning each fold change to a cluster) and "Update" (choosing a new centroid for each cluster) steps. After each "Update" step, the total cost $TC$ is calculated based on the matrix $\mathcal{OWD}$:
\begin{equation}
	TC =  \sum_{k=1}^K  \sum_{i \in cluster_k} \mathcal{OWD}_{iC_k}.
\end{equation}
The iteration continues until either the total cost stops decreasing, up to a parameter $\epsilon$, or the maximal number of iterations $it_{max}$ is reached. From the cluster configurations generated for $n_{init}$ initializations, the one with the lowest total cost is selected. Ultimately, the matrix $\mathcal{OW}$ is used to extract the final warps.

The comparison between Algorithms \ref{alg:cl_n_al_iter} and \ref{alg:cl_n_al} leads to the following result:

\begin{The}\label{the:algo}
	The following is true about the joint clustering and alignment algorithms:
	\begin{enumerate}
		\item Algorithm \ref{alg:cl_n_al} converges in a finite number of iterations.
		\item Algorithms \ref{alg:cl_n_al_iter} and \ref{alg:cl_n_al} are equivalent, in the sense that for the same input they produce the same output.
		\item Algorithms \ref{alg:cl_n_al_iter} and \ref{alg:cl_n_al} have polynomial time complexities, that are given in the proof. Moreover, the degree of the largest polynomials of the time complexity of Algorithm \ref{alg:cl_n_al_iter} is greater than that of Algorithm \ref{alg:cl_n_al}, meaning that the latter is less complex.
	\end{enumerate}
\end{The}

\begin{Rem}
	In practice, the improvement of Algorithm \ref{alg:cl_n_al} in terms of the runtime can be very important, as a large value is often required for $n_{init}$. Given the sensitivity of clustering algorithms to initialisation, it is advisable to perform a sufficient number of random initialisations to cover a sufficiently wide range of initial combinations. This is particularly important when working with higher values of $n_e$. 
\end{Rem}
\begin{Rem}
	In line \ref{lst:warps} of Algorithm \ref{alg:cl_n_al}, the indexing order of $\mathcal{OW}$ is important, since this matrix is anti-symmetric, as shown in Proposition \ref{prop:diss}. This specific indexing implies that the fold changes are being warped with respect to their centroids, which remain fixed. 
\end{Rem}

\subsection{Sign penalty}\label{sec:pen}
The following feature is driven by the realization that there is a significant distinction in interpretation between positively and negatively expressed fold changes. To reinforce this distinction in the obtained clusters, we introduce a penalty term that increases the dissimilarity for those pairs of entities with different signs for one or more corresponding instances. For a warp step $s$ and entity index pair $(i,i') \in \{1,\dots,n_e\}^2$, the penalty term represents the proportion of time points where the means of the two considered fold changes have opposite signs:

$$
Pen\left(\widehat{\Gamma_i ^{\mathcal{W}_s} }, \widehat{\Gamma_{i'} ^{\mathcal{W}_s} }\right) = \frac{1}{p-|s|} \sum_{l=1}^{p-|s|} \mathbbm{1}_{\mathbbm{R}_{-}}\left((\Gamma_{i} ^{\mathcal{W}_s} )_{t_l} \times (\Gamma_{i'} ^{\mathcal{W}_s})_{t_l}\right).
$$

By analogy with the distance matrix, a penalty matrix can be formulated:

$$
\left[ Pen_{ii'} \right]_{1 \leq i, i' \leq n_e} \text{ such that } Pen_{ii'} = Pen\left(\widehat{\Gamma_i ^{\mathcal{W}_s} }, \widehat{\Gamma_{i'} ^{\mathcal{W}_s} }\right).
$$

Finally, the penalized dissimilarity is defined with a penalization hyperparameter $\lambda \geq 0$:
$$
Pen\left( d \left(\widehat{\Gamma_i ^{\mathcal{W}_s} }, \widehat{\Gamma_{i'} ^{\mathcal{W}_s} }\right) \right) = \mathbf{\widehat{diss}_s} \left(\widehat{\Gamma_i ^{\mathcal{W}_s} }, \widehat{\Gamma_{i'} ^{\mathcal{W}_s} }\right) + \lambda \times Pen_{ii'}.
$$

This penalized dissimilarity is integrated into the aligned clustering procedure by replacing $\mathbf{\widehat{diss}_s} \left(\widehat{\Gamma_i ^{\mathcal{W}_s} }, \widehat{\Gamma_{i'} ^{\mathcal{W}_s} }\right)$ in Definitions \ref{def:owd} and \ref{def:ow}.

\section{Simulation study}\label{sec:sim}
A series of simulation studies was conducted to assess the proposed methodology in a context similar to that of temporal fold changes, in comparison to existing state-of-the-art alternatives. In each scenario, we simulated $n_e=300$ fold changes over $p=8$ time points, ranging from 0.5 to 21. Simulated fold changes are defined by their means and their covariance matrix.  In keeping with the previous notation used in the context of real datasets, the means are represented by $\Gamma = \left(\Gamma_1, \dotso, \Gamma_{n_e}\right)$, where $\Gamma_i = \left(\Gamma_{i,t_1}, \dotso, \Gamma_{i,t_p}\right)$ for $i \in \{1,  \dotso,  n_e\}$. The covariance matrices will be denoted by $\Psi  = (\Psi_{ii'})_{(i,i')\in \{1,  \dotso,  n_e\}^2}$, where $\Psi _{ii'}$ is a diagonal matrix, with the diagonal $(\psi_{\Gamma_i \Gamma_{i',t_1}}, \dots, \psi_{\Gamma_i \Gamma_{i',t_p}})$, such that, for $t \in \{t_1,  \dotso,  t_p\}$, $\psi_{\Gamma_i \Gamma_{i',t}} = \sigma^2_{\Gamma_{i}^{t}}$ if $i=i'$, and $\psi_{\Gamma_i \Gamma_{i',t}} =\rho_{\Gamma_{i,t} \Gamma_{i',t}}$ otherwise. The differences in the design of different scenarios are presented in Table \ref{tab:sim_scenarios}.
\begin{footnotesize}
				\renewcommand{\arraystretch}{0.9} 
	\linespread{0.9}
	\begin{longtable}[c]{|c|c|c|c|c|}
		\cline{2-5}
		\multicolumn{1}{c|}{} & \begin{tabular}{@{}c@{}}\textbf{Nb. of} \\[-10pt] \textbf{clusters} \end{tabular} & \multicolumn{1}{c|}{\textbf{$\Gamma_{i}$ distribution} } &  \multicolumn{1}{c|}{\textbf{$\Psi'_{ii'}$ distribution} } &  \multicolumn{1}{c|}{\textbf{$f_{\Psi}$} } \\
		
		\hline \textbf{M1-C1} & 4  & $\mathcal{U}(\{f_1, f_2, f_3, f_4\})$ & {$\!\begin{aligned}  \mathcal{N}(0, 2^2)
				\text{ if } i=i', 0 \text{ otherwise} \end{aligned}$}  & $f_{\Psi} = id$ \\
		\hline \textbf{M1-C2}  & 2  & $\mathcal{U}(\{f_1, f_2\})$ & {$\!\begin{aligned}  \mathcal{N}(0, 2^2)
				\text{ if } i=i', 0 \text{ otherwise} \end{aligned}$} & $f_{\Psi} = id$ \\
		\hline \textbf{M1-C3}  & 2  & $\mathcal{U}(\{f_1, f_2\})$ & {$\!\begin{aligned}  \lvert  \mathcal{N}(0, 2^2)  \rvert &
				\text{ if $\mathbbm{1}_{cl_1}(i)=\mathbbm{1}_{cl_1}(i')$}, \\[-3pt]  &0 \text{ otherwise} \end{aligned}$}   & $f_{\Psi}(x) = \tfrac{x^2}{cst_1}$ \\
		\hline \textbf{M1-C4}  & 2  & $\mathcal{U}(\{f_1, f_2\})$ & {$\!\begin{aligned}  \lvert  \mathcal{N}(0, 2^2)  \rvert &
				\text{ if $\mathbbm{1}_{cl_1}(i)=\mathbbm{1}_{cl_1}(i')$}, \\[-3pt]  &0 \text{ otherwise} \end{aligned}$}   & $f_{\Psi}(x) = \tfrac{x^2}{cst_2}$ \\
		\hline \textbf{M1-C5}  & 2  & $\mathcal{U}(\{f_1, f_2\})$ & {$\!\begin{aligned}  \mathcal{U}([0, 1]) \text{ if } \mathbbm{1}&_{cl_1}(i)=\mathbbm{1}_{cl_1}(i')=1, \mathcal{U}([-1, 0]) \\[-7pt] 
				\text{ if } \mathbbm{1}_{cl_2}(i)&=\mathbbm{1}_{cl_2}(i')=1,
				0 \text{ otherwise}  \end{aligned}$}   & $f_{\Psi}(x) = \tfrac{x^2}{cst_3}$ \\
		\hline \textbf{M1-C6}  & 2  & $\mathcal{U}(\{f_1, f_2\})$ & {$\!\begin{aligned}  \mathcal{U}([0, 1]) \text{ if } \mathbbm{1}&_{cl_1}(i)=\mathbbm{1}_{cl_1}(i')=1, \mathcal{U}([-1, 0]) \\[-7pt] 
				\text{ if } \mathbbm{1}_{cl_2}(i)&=\mathbbm{1}_{cl_2}(i')=1,
				0 \text{ otherwise}  \end{aligned}$}   & $f_{\Psi}(x) = \tfrac{x^2}{cst_4}$ \\
		\hline \textbf{M2}  & 4  & $\mathcal{U}(\{f_1, f_2, f_3, f_4\})$ & {$\!\begin{aligned}  \mathcal{N}(0, 2^2)
				\text{ if } i=i', 0 \text{ otherwise} \end{aligned}$}  & $f_{\Psi} = id$ \\
		\hline 
		\caption{\label{tab:sim_scenarios}Details on the simulation design in all scenarios, denoted by codes, such that M1 and M2 indicate two different approaches to the simulation of means, and C1-C6 indicate different ways to simulate covariances. The covariances are defined through a matrix and a scaling transformation: $\Psi _{ii'} = f_{\Psi}(\Psi' _{ii'})$. The scaling constants are set as follows: $\{cst_1, cst_2, cst_3, cst_4\}=\{\max \{\psi'_{\Gamma_i \Gamma_{i',t}} \vert i\neq i', t \in \{t_1, \dots, t_p\}\}, 20, 100, 50\}$.}
	\end{longtable} 
\end{footnotesize}

The design of the fold change means is detailed in Table \ref{tab:sim_means}. It should be noted that the means were simulated based on four behavior types, designed to reproduce the characteristics of the fold changes observed in the real data. The simulations are distinguished by a minimal degree of model-imposed features. Specifically, we assume that the fold changes have estimated probability distributions described by means and covariances. Thus, we simulate the fold changes estimators directly. The functional patterns of the data are not contingent on the proposed framework; they are merely inspired by those observed in the real data.

\begin{small}
	\begin{longtable}[c]{|c|c|c|}
		\cline{2-3}
		\multicolumn{1}{c|}{} & \multicolumn{1}{c|}{\textbf{M1} }& \multicolumn{1}{c|}{\textbf{M2}}\\ 
		\hline \textbf{Cluster 1}    &  $f_1(x)=\tfrac{a}{2}x^2  + bx  + c$ &  $f_1(x)= \tfrac{a}{2}(x-s)^2  + b(x-s)  + c$ \\
		\hline \textbf{Cluster 2}   & {$\!\begin{aligned}f_2(x)=&\tfrac{a}{3}x^3 - \tfrac{a(r_1+r_2)}{2}x^2 \\[-10pt] &+ (ar_1r_2+c)x  + d\end{aligned}$}   & {$\!\begin{aligned}f_2(x)=&\tfrac{a}{3}(x-s)^3 - \tfrac{a(r_1+r_2)}{2}(x-s)^2 \\[-10pt] &+ (ar_1r_2+c)(x-s)  + d\end{aligned}$}   \\
		\hline \textbf{Cluster 3}   & {$\!\begin{aligned}f_3(x)=&\tfrac{a}{3}x^3 - \tfrac{a(r_1+r_2)}{2}x^2 \\[-10pt] &+ (ar_1r_2+c)x  + d\end{aligned}$}   &  {$\!\begin{aligned}f_3(x)=&\tfrac{a}{3}(x-s)^3 - \tfrac{a(r_1+r_2)}{2}(x-s)^2 \\[-10pt] &+ (ar_1r_2+c)(x-s)  + d\end{aligned}$}  \\
		\hline \textbf{Cluster 4}   & {$\!\begin{aligned} f_4(x)=&\tfrac{a}{4}x^4 - \tfrac{a(r_3+r_4+r_5)}{3}x^3 \\[-7pt] &+ \tfrac{a\left(r_3r_4+r_5(r_1+r_2)\right)}{2}x^2 \\[-7pt] &- (ar_3r_4r_5 + b)x + c \end{aligned}$} & $f_4(x)=a \sin(b(x-s)) + c$  \\
		\hline 
		\caption{\label{tab:sim_means}Distributions of the simulated fold change means for each of the 4 behavior types (clusters) in 2 different scenarios. Each simulated fold change is an independent realization of one of these templates, with all coefficients being randomly generated (Gaussian or uniform). In particular, $s$ is the time shift parameter appearing only in the second simulation study M2, with the distribution $s \sim  \mathcal{U}([-7, 7])$. A continuous distribution was selected because, in the real world, time shifts can be reasonably assumed to be continuous. This is not something that can be captured since we only have access to measurements at discrete time points. Most parameters are used to control the shape of the fold changes to make sure each cluster represents a distinct behavior type: parameters $r_1, \dots, r_5$ are used to control the extrema of the polynomials, and the others to control the scale. See Section \ref{sec:sim_params_1} of Supplementary Material for details on the parameter distributions.}
	\end{longtable} 
\end{small}

\begin{figure}
	\hspace{-50pt}
	\includegraphics[clip, trim=5cm 0.8cm 5cm 0.5cm,scale=0.34]{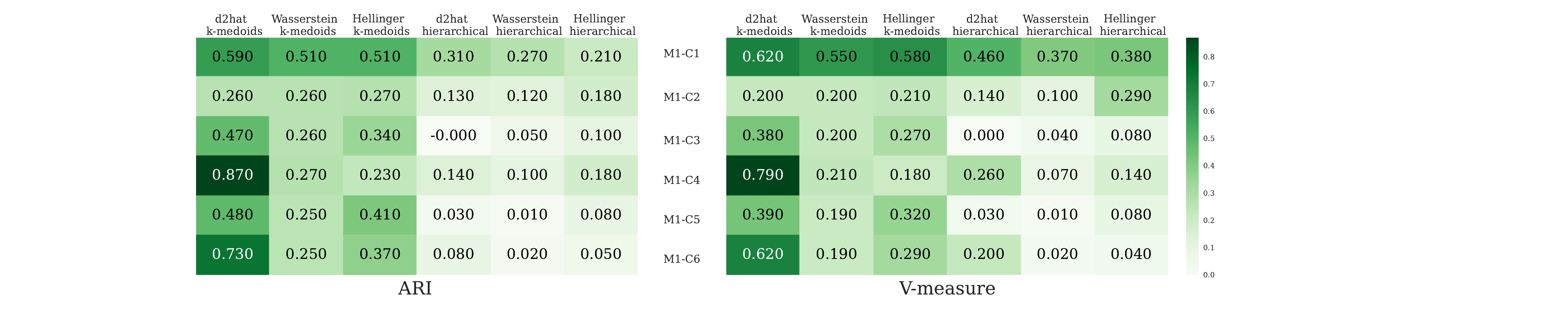}
	\caption{\label{fig:sim_cc}Results of the simulation study on distances and clustering algorithms. Mean values of the ARI \citep{chacon_minimum_2023} and V-measure \citep{rosenberg_v-measure_2007} scores are presented, standard deviation is negligible for all the methods ($\leq 10^{-15}$).}
\end{figure}

The first simulation study (scenarios M1-C1 to M1-C6) focuses on the evaluating different distances and clustering algorithms. Attempting to mimic the real data as closely as possible, the time points were spaced in the same way as in of of the real omics datasets: $(t_1,  \dotso,  t_p)=(0.5,  1,  2,  3,  4,  7,  14,  21)$. As alternatives to the $L^2$-distance, Wasserstein \citep{givens_class_1984} and Hellinger \citep{pardo_statistical_2005} distances are considered. Hierarchical clustering, implemented in Python package \textit{scikit-learn} \citep{pedregosa_scikit-learn_2011} is considered as an alternative to k-medoids \footnote{See Section \ref{sec:sim_params_2} of Supplementary Material for details on the parameters used in different methods.}. All the considered distances are implemented in \textit{scanfc}. The results for k-medoids were obtained from 50 random initializations (inapplicable to hierarchical clustering which is deterministic by nature). The study validated the effectiveness of performing $\mathbf{\widehat{d_2^2}}$-based k-medoids on fold changes-like data (Figure \ref{fig:sim_cc}). The superior performance of k-medoids compared to hierarchical clustering in independent scenarios M1-C1 and M1-C2 indicates that k-medoids is a more suitable method for clustering fold changes. The remaining scenarios, which incorporate different levels of intra-cluster covariances, demonstrate that the $\mathbf{\widehat{d_2^2}}$ distance is more effective than other distances that do not take joint distributions into account. The Hellinger distance only compares marginal distributions, while the Wasserstein distance calculates cross-covariances corresponding to the optimal mapping between the marginal distributions.

The second simulation study (scenario M2) is aimed at evaluating the effect of integrating time warping into clustering. The time points were set almost equally spaced to address the important changes in scale resulting from the horizontal shifts of the fold changes applied to study the effect of alignment:  $(t_1,  \dotso,  t_p)=(0.5,  3,  6, 9, 12,  15,  18,  21)$.

The results of this simulation study are presented in Table \ref{tab:sim2}. We compared the proposed algorithm performing k-medoids clustering jointly with alignment to k-medoids clustering without alignment, specifically applied to the distance matrix containing observed distances instead of post-warped ones. Additionally, we tested spectral clustering with and without alignment, applied to the dissimilarity matrices transformed into similarity matrices using the transformation $f(\mathcal{M})=\tfrac{\max \mathcal{M} - \mathcal{M}}{\max \mathcal{M}}$ for a given matrix $\mathcal{M}$. The spectral clustering algorithm \citep{von_luxburg_tutorial_2007} was performed using the framework implemented in Python package \textit{scikit-learn}, which consists in performing k-means clustering on relevant eigenvectors of a normalized Laplacian matrix of the similarity matrix. Finally, we tested two state-of-the-art approaches on the simulated fold change means\footnote{Both methods could only be applied to the means as they were not designed to account for replicates. The inability of these methods to take into account the correlations between entities was disregarded in order to compare other aspects.}: k-medoids clustering applied to the "distance" matrix obtained with dynamic time warping implemented in the Python package \textit{dtaidistance} \citep{meert_dtaidistance_2020}, as an alternative to our approach to alignment, and the functional version of joint clustering with alignment implemented in the R package \textit{fdacluster} \citep{stamm_fdacluster_2023}. The results for all methods were obtained from 50 random initializations, with the exception of \textit{fdacluster}, due to the unavailability of the option to modify the number of random initializations in this implementation.

\begin{footnotesize}
	\setlength\tabcolsep{5pt}
	\setlength\extrarowheight{-1.5pt}
	\begin{longtable}[c]{|c|c|c|c|c|c|c|}
		\cline{2-7}
		\multicolumn{1}{c|}{} & \multicolumn{2}{c|}{\textbf{Without alignment} }& \multicolumn{4}{c|}{\textbf{With alignment}}\\ 
		\cline{2-7}
		\multicolumn{1}{c|}{} 
		&
		\begin{tabular}{@{}c@{}}\textbf{$\mathcal{D}$} \\[-10pt] \textbf{k-medoids} \end{tabular}
		&
		\begin{tabular}{@{}c@{}}\textbf{$\mathcal{D}$ spectral} \\[-10pt] \textbf{clustering} \end{tabular} 
		&
		\begin{tabular}{@{}c@{}}\textbf{$\mathcal{OWD}$} \\[-10pt] \textbf{k-medoids} \end{tabular}
		&
		\begin{tabular}{@{}c@{}}\textbf{$\mathcal{OWD}$ spectral } \\[-10pt] \textbf{clustering} \end{tabular} 
		&
		\begin{tabular}{@{}c@{}}\textbf{DTW} \end{tabular} 
		&
		\begin{tabular}{@{}c@{}}\textit{fdacluster} \end{tabular} 
		\\
		
		\hline \textbf{ARI} & 0.22 & 0.26  & 0.61 & 0.26 & 0.30 & 0.46 \\
		\hline \textbf{V-measure} & 0.39 & 0.45 & 0.67 & 0.47 & 0.43 & 0.5 \\
		\hline
		\caption{\label{tab:sim2} Results of the simulation study of the effect of incorporating time warping into clustering. Mean values of the ARI and V-measure scores are presented, standard deviation is between 0.01 and 0.01 for the $\mathcal{D}$-based k-medoids and for dynamic time warping, and of order $10^{-3}$ and less for the rest of the methods.}
	\end{longtable} 
\end{footnotesize}

On the one hand, the study demonstrated the interest of integrating time warping to account for temporal shifts. On the other hand, it showed that our approach to joint clustering with alignment is more suitable for the data in question than various state-of-the-art alternatives. Using the example of spectral clustering, often preferred to k-medoids, we showed that the k-medoids-based clustering framework proposed in this paper leverages the information on the alignments for clustering better than other state-of-the-art approaches applied on the same dissimilarity matrix $\mathcal{OWD}$. It can also be concluded that the dynamic time warping is less adapted in this context than our approach to alignment, most likely due to the small number of time points. Similarly, our method outperforms the functional approach, while having a number of advantages with respect to the data considered in this work, such as the ability to account for correlations and lower computational time.

\section{Application to real data}\label{sec:applic}
We demonstrate the proposed approach on real data from an in vitro study taking place in the radiobiological  context. The goal is to assess the effect of irradiation with higher energy levels on human endothelial cells (HUVEC) through their transcriptomic expression. This study is part of a research project initiated at the French Radioprotection and Nuclear Safety Institute (IRSN) with the goal of improving our understanding of the adverse effects induced by radiotherapy on healthy tissues, with potential applications in a clinical setting. The transcriptomic fold changes of 157\footnote{Out of 172 measured genes, 15 were not considered due to the missingness of replicates.} genes, measured at 2, 4, 7, 14 and 21 days after irradiation, were estimated based on three replicates from the log-transformed data according to the procedure presented in Section \ref{sec:estim}. Additionally, we preprocessed the fold changes in order to amplify more significant values and reduce the difference in scale effect, the details are presented in Supplementary Material (Section \ref{sec:preproc}). We performed joint clustering with alignment based on the sign-penalized optimal warping dissimilarity matrix.

\begin{figure}
	\centering
	\begin{subfigure}{.5\textwidth}
		\centering
		\includegraphics[width=.99\linewidth]{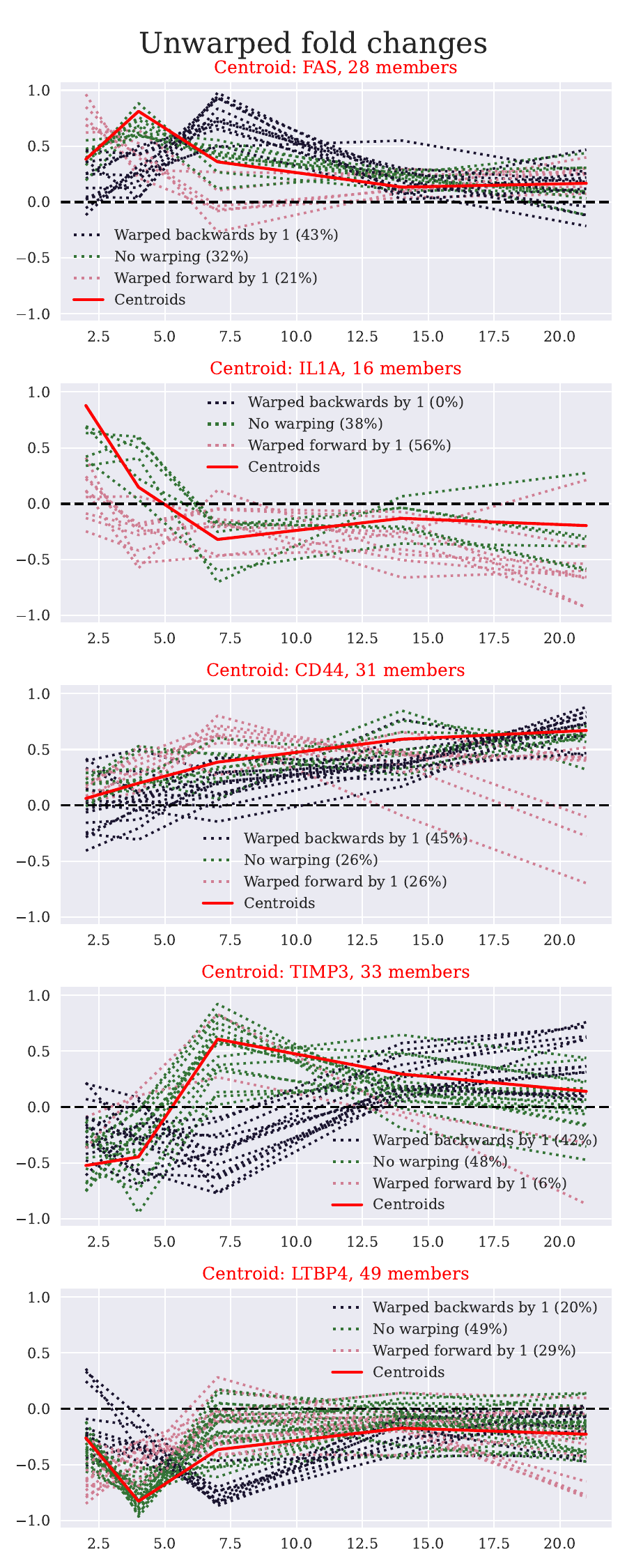}
		\caption{\label{fig:kmed_alphee_5_ua}}
	\end{subfigure}%
	\begin{subfigure}{.5\textwidth}
		\centering
		\includegraphics[width=.99\linewidth]{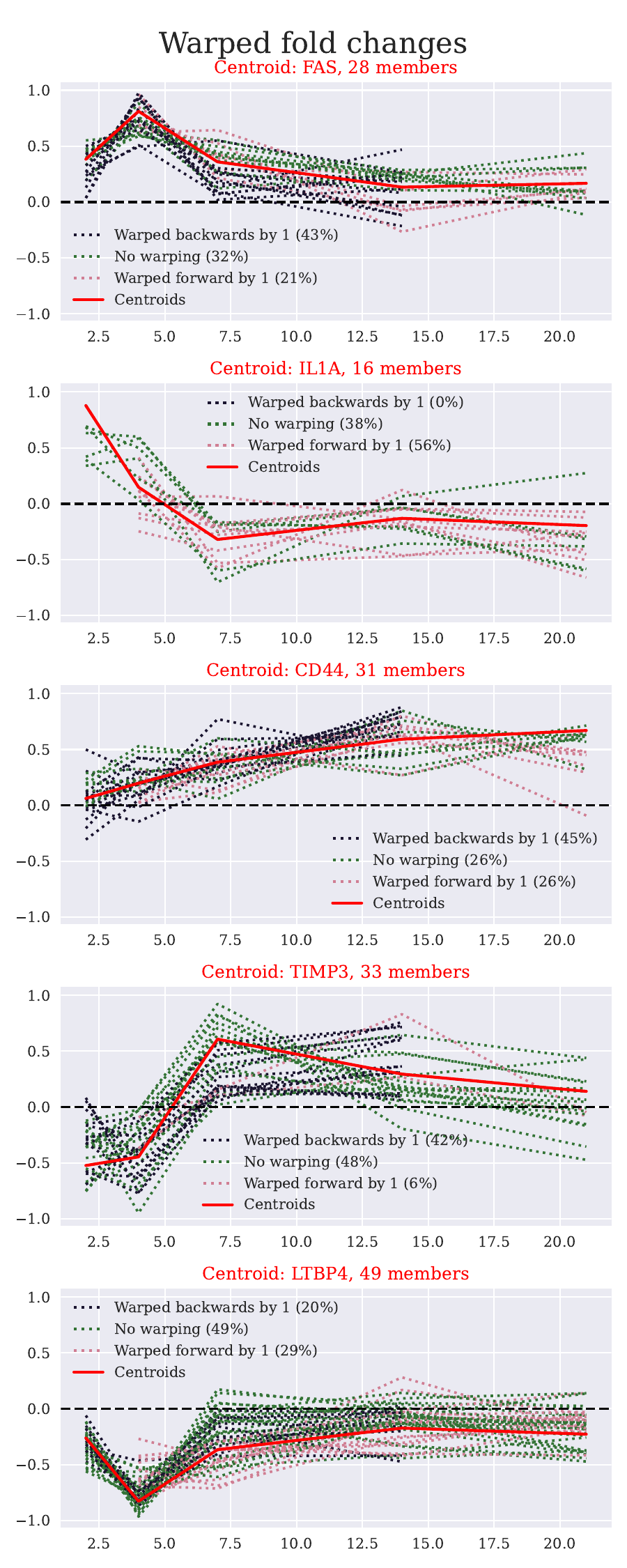}
		\caption{\label{fig:kmed_alphee_5_a}}
	\end{subfigure}
	\caption{\label{fig:kmed_alphee_5}Clustering of the LINAC dataset with $\mathcal{OWD}$-based k-medoids in 5 clusters.  a) Means of original normalized fold changes (unaligned). b) Means of warped normalized fold changes (aligned).}
\end{figure}

The five clusters\footnote{These plots only contain the means of preprocessed fold changes, giving a rough idea of the genes' behavior. However, it is important to note that they can occasionally be misleading, as clustering is performed on the full fold changes, which include not only the means but also all the information on correlations and uncertainties that can be inferred from the replicates.} that were obtained are presented in Figure \ref{fig:kmed_alphee_5}. Regarding the number of clusters, classical selection criteria such as the total cost and the silhouette score \citep{rousseeuw_silhouettes_1987}, both integrated in \textit{scanfc}, appear to favor few clusters (3 or 4, see Figure \ref{fig:cost}). However, we chose to set the number of clusters at five, as this is the smallest number that allows for a sufficient separation of features from a radiobiological perspective. The plots allow to identify which warp group (warped backward with respect to the centroid, simultaneous with the centroid and warped forward with respect to the centroid) each gene belongs to. The aligned version is the one used for clustering, allowing to identify global behavior types up to a time shift. The unaligned version allows to identify temporal cascades inside every cluster, i.e. the forward-warped predict simultaneous that predict the backward-warped. As a result, five distinct behavior types have been identified (top to bottom): up-regulated and tending towards zero, up-regulated initially and down-regulated later on, steady growth, down-regulated initially and up-regulated later on, down-regulated and tending towards zero. The clusters will be hereafter referred to by numbers according to this order. In comparison, spectral clustering\footnote{The results for both methods were obtained from 2000 random initializations.} on the similarity constructed from the $\mathcal{OWD}$ matrix, was also performed on this data in the same manner as on simulated data in the previous section. The resulting behavior types are much less distinguishable\footnote{See Supplementary Figure \ref{fig:sp_cl} for a visualization of the clusters.}, as evidenced by the silhouette score, which is 0.36 for our k-medoids-based approach and 0.24 for spectral clustering.

To gain insight into the biological interpretation of the quantities and patterns identified using our methodology, we performed an enrichment analysis of clusters with cellular processes using Pathway Studio (see Section \ref{sec:ps} of Supplementary Material for details). As a result of this analysis, for a list of genes contained in the given cluster, we obtained the cellular processes matching with an important proportion of these genes according to literature, and for each cellular process a subset of implicated genes, as well as some statistics such as the associated p-value. The processes were filtered based on the p-value at the level of 0.01 and the level of overlap, which was selected to ensure sufficient information for larger clusters while avoiding excessive detail for smaller clusters. The barplot showing the distribution of all detected cellular processes across clusters is presented in Supplementary Figure \ref{fig:EA_PS}. 

It is notable that several processes are significantly represented across multiple clusters. For instance, cell proliferation and adhesion are prominent examples. We are particularly interested in those that are highly represented in only one cluster and not the others. This allows us to conclude that the cellular process in question is likely to be a defining characteristic of this cluster. The identified cluster-specific cellular processes align with the anticipated biological outcomes. The up-regulated clusters 1 and 3 are associated with cellular aging and oxidative stress, respectively. Indeed, both functions are related to senescence/apoptosis \citep{pole_oxidative_2016}, and the implicated genes are expected to be significantly up-regulated under high-energy irradiation. Furthermore, 80\% of genes from cluster 1 that were identified as associated with cellular aging are either forward-warped or simultaneous, corresponding to an early expression pattern. This indicates the additional predictive potential of warp groups identified within given clusters. In contrast, the genes from cluster 3 that have been linked to oxidative stress demonstrate consistent growth in expression, reaching a peak in the late phase. This may have been influenced by the aforementioned genes from cluster 1.

The genes from cluster 5, which has a down-regulated pattern symmetrically to cluster 1, are associated with cell survival. This function is of an opposing nature with respect to apoptosis, and the corresponding genes are expected to show strong down-regulation in this context. The remaining clusters are more difficult to interpret due to their mixed nature. Cluster 2 is the smallest in size and appears to primarily contain genes with endothelium-specific functions. Lastly, cluster 4 has been linked to the process of DNA replication. The observed pattern of early down-regulation with a subsequent up-regulation of these genes is consistent with the hypothesis that the DNA replication process initially slows down as a reaction to stress, and is then reinforced further on in order to restore the damaged tissues.

\section{Discussion}
In this article we propose a novel procedure designed for complex temporal data in the context where only one time point can be observed per experiment. We designed a computationally efficient algorithm performing clustering jointly with temporal alignment of the fold changes, chosen as the object encoding the response in the context of two different treatment groups. Our approach to modeling the fold changes, the proposed dissimilarity measure, and the main algorithm have been shown to be more adapted for the treated setting than several alternative options. Moreover, our method was successfully applied to real transcriptomic data, enabling the identification of biologically meaningful groups of genes with interpretable patterns. The proposed approach could potentially be extended to a more sophisticated form of clustering, which may produce even better results on complex data.

\section*{Software availability}
Python package and the LINAC data are freely available at \href{https://github.com/parsenteva/scanfc}{https://github.com/parsenteva/scanfc}.

\section*{Acknowledgments}
We thank Olivier Guipaud, Fabien Milliat and Vincent Paget for the experimental data and useful discussions. This work is supported by the European Union through the PO FEDER-FSE Bourgogne 2014/2020 programs as part of the project ModBioCan2020, and by Institut de Radioprotection et de Sûreté nucléaire as part of the project ROSIRIS.

\textit{Conflict of Interest:} none declared.

\appendix

\counterwithin{figure}{section}
\counterwithin{table}{section}
\counterwithin{equation}{section}

\section{Proofs}
\begin{Proof}\textbf{ of Proposition \ref{prop:diss}} \\
	We will denote the joint distribution of the fold changes pair $\left[ \widehat{\Gamma_{i} ^{\mathcal{W}_s} } ^\intercal  \widehat{\Gamma_{i'} ^{\mathcal{W}_s} } ^\intercal \right]^\intercal$ in the following way:
	$$
	\begin{bmatrix} \widehat{\Gamma_i ^{\mathcal{W}_s} } \\ \widehat{\Gamma_{i'} ^{\mathcal{W}_s} }  \end{bmatrix} \sim \mathcal{N}\left(\begin{bmatrix} \Gamma_i ^{\mathcal{W}_s} \\ \Gamma_{i'} ^{\mathcal{W}_s}  \end{bmatrix}, \begin{bmatrix} 
		\Sigma_{\Gamma_i} ^{\mathcal{W}_s}  & \mathrm{P}_{\Gamma_i \Gamma_{i'}}  ^{\mathcal{W}_s}  \\
		(\mathrm{P}_{\Gamma_i \Gamma_{i'}}  ^{\mathcal{W}_s}) ^\intercal & \Sigma_{\Gamma_{i'}} ^{\mathcal{W}_s} 
	\end{bmatrix} \right).
	$$
	
	Using Definitions \ref{def:warped_fc_pair} and \ref{def:fc_pair}, the means can be expressed depending on the warp type:
	$$
	\begin{bmatrix} \Gamma_i ^{\mathcal{W}_s} \\ \Gamma_{i'} ^{\mathcal{W}_s}  \end{bmatrix} = \begin{cases}
		\left[ \Gamma_{i,t_{1}} \hdots \Gamma_{i,t_{p-|s|}}  \Gamma_{i',t_{1+|s|}}  \hdots \Gamma_{i',t_{p}} \right]^\intercal \text{ if } s>0 \\
		\left[ \Gamma_{i,t_{1+|s|}} \hdots \Gamma_{i,t_{p}}  \Gamma_{i',t_{1}} \hdots \Gamma_{i',t_{p-|s|}}  \right]^\intercal \text{ if } s<0 \\
		\left[ \Gamma_{i}^\intercal \Gamma_{i'}^\intercal \right]^\intercal \text{ if } s=0
	\end{cases},
	$$
	
	Similarly, we can express the elements of the covariance matrix:
	$$
	\Sigma_{\Gamma_i} ^{\mathcal{W}_s} = \begin{cases}
		\begin{bmatrix} \sigma^2_{\Gamma_{i,t_1}} & & 0 \\ & \ddots \\ 0 & & \sigma^2_{\Gamma_{i,t_{p-|s|}}}  \end{bmatrix} \text{ if } s>0 \\[40pt]
		\begin{bmatrix} \sigma^2_{\Gamma_{i,t_{1+|s|}}} & & 0 \\ & \ddots \\ 0 & & \sigma^2_{\Gamma_{i,t_{p}}}  \end{bmatrix} \text{ if } s<0 \\[40pt]
		\Sigma_{\Gamma_i} \text{ if } s=0
	\end{cases}, \hspace{15pt}
	\Sigma_{\Gamma_{i'}} ^{\mathcal{W}_s} = \begin{cases}
		\begin{bmatrix} \sigma^2_{\Gamma_{i',t_{1+|s|}}} & & 0 \\ & \ddots \\ 0 & & \sigma^2_{\Gamma_{i',t_{p}}}  \end{bmatrix} \text{ if } s>0 \\[40pt]
		\begin{bmatrix} \sigma^2_{\Gamma_{i',t_{1}}} & & 0 \\ & \ddots \\ 0 & & \sigma^2_{\Gamma_{i',t_{p-|s|}}}  \end{bmatrix}\text{ if } s<0 \\[30pt]
		\Sigma_{\Gamma_{i'}} \text{ if } s=0
	\end{cases}.
	$$
	\vspace{0.5cm}
	$$
	\text{ and } \hspace{15pt}
	\mathrm{P}_{\Gamma_i \Gamma_{i'}} ^{\mathcal{W}_s} = \begin{cases}
		\begin{tikzpicture}[baseline=(current bounding box.center)]
			\fontsize{8}{10}\selectfont
			\matrix (m) [matrix of math nodes,nodes in empty cells,right delimiter={]},left delimiter={[} ]{
				0 &  &  &  & 0  \\ 
				&  &  &  &    \\
				\rho_{\Gamma_{i,t_{1+|s|}}\Gamma_{i',t_{1+|s|}}} &   & & &  \\
				&  &  &  &   \\
				0 & & \rho_{\Gamma_{i,t_{p-|s|}}\Gamma_{i',t_{p-|s|}}} &  &  0 \\ 
			} ;
			\draw[loosely dotted] (m-1-1)-- (m-5-5);
			\draw[loosely dotted] (m-1-1)-- (m-3-1);
			\draw[loosely dotted] (m-3-1)-- (m-5-3);
			\draw[loosely dotted] (m-5-3)-- (m-5-5);
		\end{tikzpicture} \text{ if } s>0 \\[35pt]
		\begin{tikzpicture}[baseline=(current bounding box.center)]
			\fontsize{8}{10}\selectfont
			\matrix (m) [matrix of math nodes,nodes in empty cells,right delimiter={]},left delimiter={[} ]{
				0 &  & \rho_{\Gamma_{i,t_{1+|s|}}\Gamma_{i',t_{1+|s|}}}  &  & 0  \\ 
				&  &  &  &    \\
				&   & & & \rho_{\Gamma_{i,t_{p-|s|}}\Gamma_{i',t_{p-|s|}}} \\
				&  &  &  &   \\
				0 & & &  &  0 \\ 
			} ;
			\draw[loosely dotted] (m-1-1)-- (m-5-5);
			\draw[loosely dotted] (m-1-1)-- (m-1-3);
			\draw[loosely dotted] (m-1-3)-- (m-3-5);
			\draw[loosely dotted] (m-3-5)-- (m-5-5);
		\end{tikzpicture} \text{ if } s<0 \\[35pt]
		\Sigma_{\Gamma_{i'}} \text{ if } s=0
	\end{cases}
	$$
	\vspace{0.7cm}
	
	It can be noted that since the non-zero elements of the matrix $K^{\mathcal{W}_s}$ have been moved from the diagonal to either sub-diagonal or super-diagonal of order $s$, its trace is now equal to zero. Since the condition on the joint distribution given in Definition \ref{def:diss} is satisfied, we can calculate the value of $\mathbf{\widehat{diss}_s}$ element by element, starting with the square norm of the difference between means:
	
	$$
	\Vert \Gamma_i ^{\mathcal{W}_s} -  \Gamma_{i'} ^{\mathcal{W}_s} \Vert ^2 = \begin{cases}
		\sum_{l=1}^{p-|s|} \left( \Gamma_{i,t_l} - \Gamma_{i',t_{l+s}} \right)^2 \text{ if } s>0 \\
		\sum_{l=1+|s|}^{p} \left( \Gamma_{i,t_l} - \Gamma_{i',t_{l+s}} \right)^2 \text{ if } s<0 \\
		\sum_{l=1}^{p} \left( \Gamma_{i,t_l} - \Gamma_{i',t_{l+s}} \right)^2 \text{ if } s=0
	\end{cases}.\
	$$
	Next, the trace of the covariance matrix of the first warped fold change in the pair:
	$$
	\mathrm{Tr} (\Sigma_ {\Gamma_i} ^{\mathcal{W}_s}) =  \begin{cases}
		\sum_{l=1}^{p-|s|} \sigma^2_{\Gamma_{i,t_l}} \text{ if } s>0 \\
		\sum_{l=1+|s|}^{p} \sigma^2_{\Gamma_{i,t_l}} \text{ if } s<0 \\
		\sum_{l=1}^{p} \sigma^2_{\Gamma_{i,t_l}}  \text{ if } s=0
	\end{cases}.\
	$$
	Similarly for the second fold change:
	$$
	\mathrm{Tr} (\Sigma_ {\Gamma_{i'}} ^{\mathcal{W}_s}) =   \begin{cases}
		\sum_{l=1+|s|}^{p} \sigma^2_{\Gamma_{i,t_l}}=\sum_{l=1}^{p-|s|} \sigma^2_{\Gamma_{i,t_{l+s}}}  \text{ if } s>0 \\
		\sum_{l=1}^{p-|s|} \sigma^2_{\Gamma_{i,t_l}}= \sum_{l=1+|s|}^{p} \sigma^2_{\Gamma_{i,t_{l+s}}} \text{ if } s<0 \\
		\sum_{l=1}^{p} \sigma^2_{\Gamma_{i,t_l}} = \sum_{l=1}^{p} \sigma^2_{\Gamma_{i,t_{l+s}}}  \text{ if } s=0
	\end{cases}.\
	$$
	Finally, the last term containing the cross-covariances is calculated:
	$$
	\sum_{l=1}^{p-|s|} \left[\mathrm{P}_{\Gamma_i \Gamma_{i'}}  ^{\mathcal{W}_s} \right]_{\left(l+s \mathbbm{1}_{\mathbbm{Z}_{+}^{*}}(s),  l+s \mathbbm{1}_{\mathbbm{Z}_{-}^{*}}(s)\right)} =   \begin{cases}
		\sum_{l=1+|s|}^{p-|s|} \rho_{\Gamma_{i,t_l} \Gamma_{i',t_l}}  \text{ if } s>0 \\
		\sum_{l=1+|s|}^{p-|s|} \rho_{\Gamma_{i,t_l} \Gamma_{i',t_l}} \text{ if } s<0 \\
		\sum_{l=1 }^{p} \rho_{\Gamma_{i,t_l} \Gamma_{i',t_l}} \text{ if } s=0
	\end{cases}.\
	$$
	Hence, we obtain the value of the dissimilarity by writing the expression for any step $s \in \mathbbm{Z}$.
\end{Proof}

\begin{Proof}\textbf{ of Proposition \ref{prop:sym}} \\
	Let $(i, i') \in \{1, \dots, n_e\}^2$ be an entity pair. The statements of the proposition are equivalent to saying that, for any warp step $s \in \mathcal{S}$, we have:
	\begin{equation}
		\begin{cases}
			\min_{s \in \mathcal{S}} \left[ \mathbf{\widehat{diss}_s}\left(\widehat{\Gamma_i ^{\mathcal{W}_s} }, \widehat{\Gamma_{i'} ^{\mathcal{W}_s} }\right) \right] = \min_{s \in \mathcal{S}} \left[ \mathbf{\widehat{diss}_s}\left(\widehat{\Gamma_{i'} ^{\mathcal{W}_s} }, \widehat{\Gamma_{i} ^{\mathcal{W}_s} }\right) \right] \\
			\arg \min_{s \in \mathcal{S}} \left[ \mathbf{\widehat{diss}_s}\left(\widehat{\Gamma_i ^{\mathcal{W}_s} }, \widehat{\Gamma_{i'} ^{\mathcal{W}_s} }\right) \right] = -\arg \min_{s \in \mathcal{S}} \left[ \mathbf{\widehat{diss}_s}\left(\widehat{\Gamma_{i'} ^{\mathcal{W}_s} }, \widehat{\Gamma_{i} ^{\mathcal{W}_s} }\right) \right] 
		\end{cases}.\
	\end{equation}
	
	Let us denote $s_*=\arg \min_{s \in \mathcal{S}} \left[ \mathbf{\widehat{diss}_s}\left(\widehat{\Gamma_i ^{\mathcal{W}_s} }, \widehat{\Gamma_{i'} ^{\mathcal{W}_s} }\right) \right]$. To prove both parts of the proposition, it suffices to show that the following is true:
	\begin{equation}\label{eq:OWD_OW}
		\mathbf{\widehat{diss}_s}\left(\widehat{\Gamma_i ^{\mathcal{W}_{s_*}} }, \widehat{\Gamma_{i'} ^{\mathcal{W}_{s_*}} }\right) = \mathbf{\widehat{diss}_s}\left(\widehat{\Gamma_{i'} ^{\mathcal{W}_{-s_*}} }, \widehat{\Gamma_{i} ^{\mathcal{W}_{-s_*}} }\right)
	\end{equation}
	
	Using the expression of the dissimilarity given in Proposition \ref{prop:diss}, we can develop the left-hand side of \eqref{eq:OWD_OW}:
	\begin{equation}\label{eq:diss_sym}
		\mathbf{\widehat{diss}_s} \left(\widehat{\Gamma_i ^{\mathcal{W}_{s_*}} }, \widehat{\Gamma_{i'} ^{\mathcal{W}_{s_*}} }\right) 
		= \sum_{l=l^{*}}^{p^{*}} \left( \Gamma_{i,t_l} - \Gamma_{i',t_{l+s_*}} \right)^2 + \sum_{l=l^{*}}^{p^{*}} \sigma^2_{\Gamma_{i,t_l}} + \sum_{l=l^{*}}^{p^{*}} \sigma^2_{\Gamma_{i',t_{l+s_*}}} - 2 \sum_{l=1+|s_*| }^{p-|s_*|} \rho_{\Gamma_{i,t_l} \Gamma_{i',t_l}}. 
	\end{equation}
	where $l^{*}=1-s_*\mathbbm{1}_{\mathbbm{Z}_{-}^{*}}(s_*)$ and $p^{*}=p-s_* \mathbbm{1}_{\mathbbm{Z}_{+}^{*}}(s_*)$.
	
	Similarly, we develop the right-hand side:
	\begin{equation}\label{eq:diss_antisym}
		\mathbf{\widehat{diss}_s}\left(\widehat{\Gamma_{i'} ^{\mathcal{W}_{-s_*}} }, \widehat{\Gamma_{i} ^{\mathcal{W}_{-s_*}} }\right)
		= \sum_{l=l_{*}}^{p_{*}} \left( \Gamma_{i',t_l} - \Gamma_{i,t_{l-s_*}} \right)^2 + \sum_{l=l_{*}}^{p_{*}} \sigma^2_{\Gamma_{i',t_l}} + \sum_{l=l_{*}}^{p_{*}} \sigma^2_{\Gamma_{i,t_{l-s_*}}} - 2 \sum_{l=1+|-s_*| }^{p-|-s_*|} \rho_{\Gamma_{i',t_l} \Gamma_{i,t_l}}. 
	\end{equation}
	where $l_{*}=1-(-s_*)\mathbbm{1}_{\mathbbm{Z}_{-}^{*}}(-s_*)$ and $p_{*}=p-(-s_*)\mathbbm{1}_{\mathbbm{Z}_{+}^{*}}(-s_*)$. These quantities can be rewritten as $l_{*}=1+s_*\mathbbm{1}_{\mathbbm{Z}_{+}^{*}}(s_*)=l^{*}+s_*$ and $p_{*}=p+s_*\mathbbm{1}_{\mathbbm{Z}_{-}^{*}}(s_*)=p^{*}+s_*$. It can also be noticed that $|-s_*|=|s_*|$, and $ \rho_{\Gamma_{i',t_l} \Gamma_{i,t_l}}= \rho_{\Gamma_{i,t_l} \Gamma_{i',t_l}}$ by the symmetry of the covariance. Hence, we can rewrite \eqref{eq:diss_antisym} as follows:
	\begin{align}
		\begin{split}\label{eq:diss_antisym_develop}
			\mathbf{\widehat{diss}_s}\left(\widehat{\Gamma_{i'} ^{\mathcal{W}_{-s_*}} }, \widehat{\Gamma_{i} ^{\mathcal{W}_{-s_*}} }\right)
			= & \sum_{l=l^{*}+s_*}^{p^{*}+s_*} \left(\Gamma_{i,t_{l-s_*}} - \Gamma_{i',t_l}  \right)^2 + \sum_{l=l^{*}+s_*}^{p^{*}+s_*} \sigma^2_{\Gamma_{i,t_{l-s_*}}}\\ &+ \sum_{l=l^{*}+s_*}^{p^{*}+s_*} \sigma^2_{\Gamma_{i',t_l}} - 2 \sum_{l=1+|s_*| }^{p-|s_*|} \rho_{\Gamma_{i,t_l} \Gamma_{i',t_l}}. 
		\end{split}
	\end{align}
	It can be noticed that the expression in \eqref{eq:diss_antisym_develop} is identical to \eqref{eq:diss_sym}, which concludes the proof.
\end{Proof}

\begin{algorithm}
	\linespread{0.9}
	\footnotesize
	\caption{Joint clustering and alignment algorithm: classical version}\label{alg:cl_n_al_iter}
	\begin{algorithmic}[1]
		\Require Fold changes  $\widehat{\Gamma}=\left(\widehat{\Gamma}_1, \dots, \widehat{\Gamma}_{n_e}\right)$, $K \in \mathbbm{N}$, $it_{max} \in \mathbbm{N}$, $n_{init} \in \mathbbm{N}$, $\epsilon>0$.
		\State $TC \gets \infty$ \Comment{global total cost} 
		\For{$init \in \{1, \dots, n_{init} \}$}
		\State Initialize centroids $C = \left(C_1 , \dots,  C_K \right) \subset  \{1 , \dots,  n_e\} $ with kmeans++
		\State $TC_{it} \gets \infty$ \Comment{total cost of the current initialization} 
		\State $\Delta TC \gets \infty$ \Comment{change in total cost of the current initialization} 
		\State $it \gets 1$
		\While{$\Delta TC > \epsilon$ \textbf{and} $it < it_{max}$}
		\State  $TC_{it}^{new} \gets 0$  
		
		\State \textbf{1. Assign step:}
		\For{$i \in\{1 , \dots,  n_e\}$}
		\State $d_{min} \gets \infty $		 \label{lst:assign_iter_first}
		\For{$k \in \{1, \dots, K \}$}
		\State $s_{k} \gets \argmin_{s \in \mathcal{S}}\mathbf{\widehat{diss}_s}\left(\widehat{\Gamma_i ^{\mathcal{W}_s}}, \widehat{\Gamma_{C_k} ^{\mathcal{W}_s}} \right)$ \Comment{Align fold changes with the centroid}
		\State $d_k \gets \mathbf{\widehat{diss}_{s_k}}\left(\widehat{\Gamma_i ^{\mathcal{W}_{s_k}}}, \widehat{\Gamma_{C_k} ^{\mathcal{W}_{s_k}}} \right)$
		\If{$d_k  < d_{min}$}  \Comment{Assign fold changes to centroids}
		\State $d_{min} \gets d_k $
		\State $Cl_i  \gets k $
		\EndIf
		\EndFor  \label{lst:assign_iter_last}
		\EndFor 
		
		\State  \textbf{2. Update step:}
		\For{$k \in \{1, \dots, K \}$} 
		\State $d_{min} \gets \infty $
		\For{$i \in cluster_k = \{ i \in \{1 , \dotso,  n_e\} | Cl_i=k\}$} \Comment{Candidate for a centroid}
		\State $d_{cluster_k} \gets 0 $  \label{lst:update_iter_first}
		\For{$i' \in cluster_k$}
		\State $s_{i'i} \gets \argmin_{s \in \mathcal{S}}\mathbf{\widehat{diss}_s}\left(\widehat{\Gamma_{i'} ^{\mathcal{W}_{s}}},  \widehat{\Gamma_{i} ^{\mathcal{W}_{s}}}  \right)$\Comment{Align fold changes with the candidate}
		\State $d_{cluster_k} \gets d_{cluster_k} +  \mathbf{\widehat{diss}_{s_{i'i}}}\left(\widehat{\Gamma_{i'} ^{\mathcal{W}_{s_{i'i}}}},  \widehat{\Gamma_{i} ^{\mathcal{W}_{s_{i'i}}}}  \right)$
		\EndFor
		\If{$d_{cluster_k} < d_{min}$} \Comment{Choose new centroid}
		\State $d_{min} \gets d_{cluster_k} $
		\State $C_k^{new}  \gets i $
		\EndIf   \label{lst:update_iter_last}
		\EndFor
		\State  $TC_{it}^{new} \gets TC_{it}^{new} + d_{min} $  
		\EndFor
		
		\State   \textbf{3. Calculate the change in total cost: } 
		\State  $\Delta TC \gets TC_{it} -  TC_{it}^{new}$ 
		\If{$\Delta TC  > \epsilon$}
		\State $C \gets \left(C_1^{new} , \dots,  C_K^{new} \right) $
		\State $TC_{it} \gets TC_{it}^{new}$
		\EndIf
		\State $it \gets it + 1$
		\EndWhile
		\If{$TC_{it} < TC$}
		\State $C \gets  \left(C_1 , \dots,  C_K \right)$\Comment{centroids labels}
		\State $Cl \gets  \left(Cl_1 , \dots,  Cl_{n_e} \right)$\Comment{cluster labels}
		\State $\mathcal{W} \gets  \left(s_{1Cl_{1}}, \dots, s_{n_e Cl_{n_e}} \right)$\Comment{warps}  \label{lst:warps_iter}
		\State $TC \gets TC_{it} $
		\EndIf
		\EndFor
		\State \Return $C$, $Cl$,  $\mathcal{W}$
	\end{algorithmic}
\end{algorithm}

\begin{Proof}\textbf{ of Theorem \ref{the:algo}} \\
	To prove the first statement, it suffices to show that the total cost always decreases, that is, for every iteration $it$, $TC_{it} \geq TC_{it+1}$.
	
	We denote $\left(Cl_1 , \dots,  Cl_{n_e} \right)$ and $\left(Cl_1^* , \dots,  Cl_{n_e}^* \right)$ cluster labels at iterations $it$ and $it+1$ respectively. For a given initialization, the cost at iteration $it$ can be written as follows:
	\begin{equation}
		TC_{it} = \sum_{k=1}^{K} \sum_{i \in cluster_k}\mathcal{OWD}_{iC_k},
	\end{equation}
	given, for every cluster label $k \in \{1, \dots, K\}$, $cluster_k = \{ i \in \{1 , \dotso,  n_e\} | Cl_i=k\}$ the current composition of the cluster, and $C_k$ the corresponding centroid. Similarly, the cost at iteration $it+1$ can be expressed:
	\begin{equation}
		TC_{it+1} = \sum_{k=1}^{K} \sum_{i \in cluster_k^*}\mathcal{OWD}_{iC_k^*},
	\end{equation}
	given, for every cluster label $k \in \{1, \dots, K\}$, $cluster_k^* = \{ i \in \{1 , \dotso,  n_e\} | Cl_i^*=k\}$ the current composition of the cluster, and $C_k^*$ the corresponding centroid. Additionally, for a given cluster label $k$, we denote the migrating sub-clusters: 
	\begin{itemize}
		\item[--] the sub-cluster of elements that left cluster $k$ at $it+1$:
		$$
		cluster_k^{*C} = \{ i \in \{1 , \dotso,  n_e\} | Cl_i=k \text{ and } Cl_i^* \neq k\},
		$$
		\item[--] the sub-cluster of elements that joined cluster $k$ at $it+1$:
		$$
		cluster_k^{C} = \{ i \in \{1 , \dotso,  n_e\} | Cl_i \neq k \text{ and } Cl_i^*=k\}.
		$$
	\end{itemize}
	
	Noticing that $cluster_k = \left( cluster_k^{*} \cup cluster_k^{*C} \right) \setminus cluster_k^{C}$, the quantity $TC_{it}$ can be decomposed as follows:
	\begin{equation}
		TC_{it} = \underbrace{\sum_{k=1}^{K} \sum_{i \in cluster_k^*}\mathcal{OWD}_{iC_k}}_\text{A} + \underbrace{\sum_{k=1}^{K} \sum_{i \in cluster_k^{*C}}\mathcal{OWD}_{iC_k}}_\text{B} - \underbrace{\sum_{k=1}^{K} \sum_{i \in cluster_k^C}\mathcal{OWD}_{iC_k}}_\text{C}.
	\end{equation}
	
	First, it follows from the "Update" step by construction that 
	$$
	A=\sum_{k=1}^{K} \sum_{i \in cluster_k^*}\mathcal{OWD}_{iC_k} \geq \sum_{k=1}^{K} \sum_{i \in cluster_k^*}\mathcal{OWD}_{iC_k^*} = TC_{it+1}.
	$$
	
	Next, we consider the quantities B and C. It can be noticed, by construction of the "Assign" step, that for every $i \in cluster_k^{*C}$ there exists a unique $k* \in \{1, \dots, K\} \setminus {k}$ such that $i \in cluster_{k*}^{C}$ and $\mathcal{OWD}_{iC_{k^*}} \leq \mathcal{OWD}_{iC_{k}}$. In other words, there is a bijection between the indices in B and C, such that the corresponding elements of the sum in B are larger than those in C. Therefore, $\text{B} -\text{C} \geq 0$, and $TC_{it}=\text{A}+\text{B} -\text{C}  \geq TC_{it+1}$.
	
	Thus, the total cost sequence is decreasing, and, noticing that total cost is positive, it can be concluded that the sequence has a limit. Finally, there is a finite number of cluster configurations possible, therefore the sequence of total costs contains a finite number of values. Hence, the algorithm converges in a finite number of iterations.
	
	The second statement follows directly from Definitions \ref{def:owd} and \ref{def:ow}. In particular, we have:
	\begin{itemize}
		\item[--] Line \ref{lst:assign} of Algorithm \ref{alg:cl_n_al} is equivalent to lines \ref{lst:assign_iter_first}-\ref{lst:assign_iter_last} of Algorithm \ref{alg:cl_n_al_iter}, since: 
		$$
		\argmin_{k \in \{1 , \dotso,  K\} } \mathcal{OWD}_{iC_k} = \argmin_{k \in \{1 , \dotso,  K\}}  \left( \min_{s \in \mathcal{S}} \left[ \mathbf{\widehat{diss}_s}\left(\widehat{\Gamma_i ^{\mathcal{W}_s} }, \widehat{\Gamma_{C_k} ^{\mathcal{W}_s} }\right) \right] \right) = 	\argmin_{k \in \{1 , \dotso,  K\} } (d_k),
		$$
		where $d_k$ is the quantity from Algorithm \ref{alg:cl_n_al_iter} of the final value after the for loop terminating at line \ref{lst:assign_iter_last}.
		
		\item[--] Lines \ref{lst:update_first}-\ref{lst:update_last} of Algorithm \ref{alg:cl_n_al} are equivalent to lines \ref{lst:update_iter_first}-\ref{lst:update_iter_last} of Algorithm \ref{alg:cl_n_al_iter}, since:
		$$
		\sum_{i' \in cluster_k}\mathcal{OWD}_{ii'} = \sum_{i' \in cluster_k} \mathcal{OWD}_{i'i} = \sum_{i' \in cluster_k} \mathbf{\widehat{diss}_s}\left(\widehat{\Gamma_{i'} ^{\mathcal{W}_s} }, \widehat{\Gamma_{i} ^{\mathcal{W}_s} }\right),
		$$
		due to the symmetry of $\mathcal{OWD}$, the final quantity being equivalent to the value of $d_{cluster_k}$ at line \ref{lst:update_iter_last} of Algorithm \ref{alg:cl_n_al_iter}. 
		\item[--] Line \ref{lst:warps} of Algorithm \ref{alg:cl_n_al} is equivalent to line \ref{lst:warps_iter} of Algorithm \ref{alg:cl_n_al_iter}.
	\end{itemize}
	We obtain the following complexities for different parts of the algorithms:
	\begin{enumerate}[font=\bfseries]
		\item \textbf{Assign step: } $\mathcal{O}(n_eK|\mathcal{S}|)$ for Algorithm \ref{alg:cl_n_al_iter}, and $\mathcal{O}(n_eK)$ for Algorithm \ref{alg:cl_n_al}.
		\item \textbf{Update step: } $\mathcal{O}(n_e^2K|\mathcal{S}|)$ for Algorithm \ref{alg:cl_n_al_iter}, and $\mathcal{O}(n_e^2K)$ for Algorithm \ref{alg:cl_n_al}.
	\end{enumerate}
	Thus, adding the complexity of calculating the matrices $\mathcal{OWD}$ and $\mathcal{OW}$ beforehand, we obtain in total $\mathcal{O}(n_{init}it_{max}n_eK|\mathcal{S}|(1+n_e))$ for Algorithm \ref{alg:cl_n_al_iter}, and $\mathcal{O}(n_{init}it_{max}n_eK(1+n_e)+n_e^2|\mathcal{S}|)$ for Algorithm \ref{alg:cl_n_al}. The degree of the largest polynomial of the former is 6, and that of the latter is 5, hence Algorithm \ref{alg:cl_n_al} is less complex.
\end{Proof}

\section{Data preprocessing}\label{sec:preproc}
Before performing clustering, certain transformations have to be applied to the raw data in order to amplify those characteristics that are of particular interest, and reduce those that can be ignored.  We perform data scaling with respect to the following criteria:
\begin{itemize}
	\item \textbf{Scaling by standard deviation:} performed in order to account for uncertainties, so that the observations with high uncertainty caused by individual variability appear with lower weight compared to those with low uncertainty. Standard deviation estimates are calculated as follows: for $i \in \{1,2, \dotso,n_e\}$ we denote $\sigma_{\Gamma_i} = (\sigma_{\Gamma_{i,t_1}}, \dotso, \sigma_{\Gamma_{i,t_p}})$ where $\sigma_{\Gamma_{i,t}} = \sqrt{\sigma^2_{\Gamma_{i,t}}}$.
	\item \textbf{Scaling by the fold change norm:} performed with the purpose of diminishing the effect of scale differences between the fold changes. The norm of $\widehat{\Gamma}_i$ associated with the distance $\mathbf{\widehat{d_2^2}}$ or the dissimilarity $\mathbf{\widehat{diss}_s}$ can be expressed as follows:
	\begin{equation}\label{eq:fc_norm}
		Norm(\widehat{\Gamma}_i) = \sqrt{\Vert \Gamma_i \Vert _2^2 + \mathrm{Tr}(\Sigma_{\Gamma_i})} = \sqrt{\Vert  \Gamma_i \Vert _2^2 + \sum_{l=1}^{p}\sigma^2_{\Gamma_{i,t_l}}}.
	\end{equation}
\end{itemize}

\begin{figure}[H]
	\centering
	\includegraphics[clip, trim=2cm 0.5cm 2cm 0.5cm,width=0.9\textwidth]{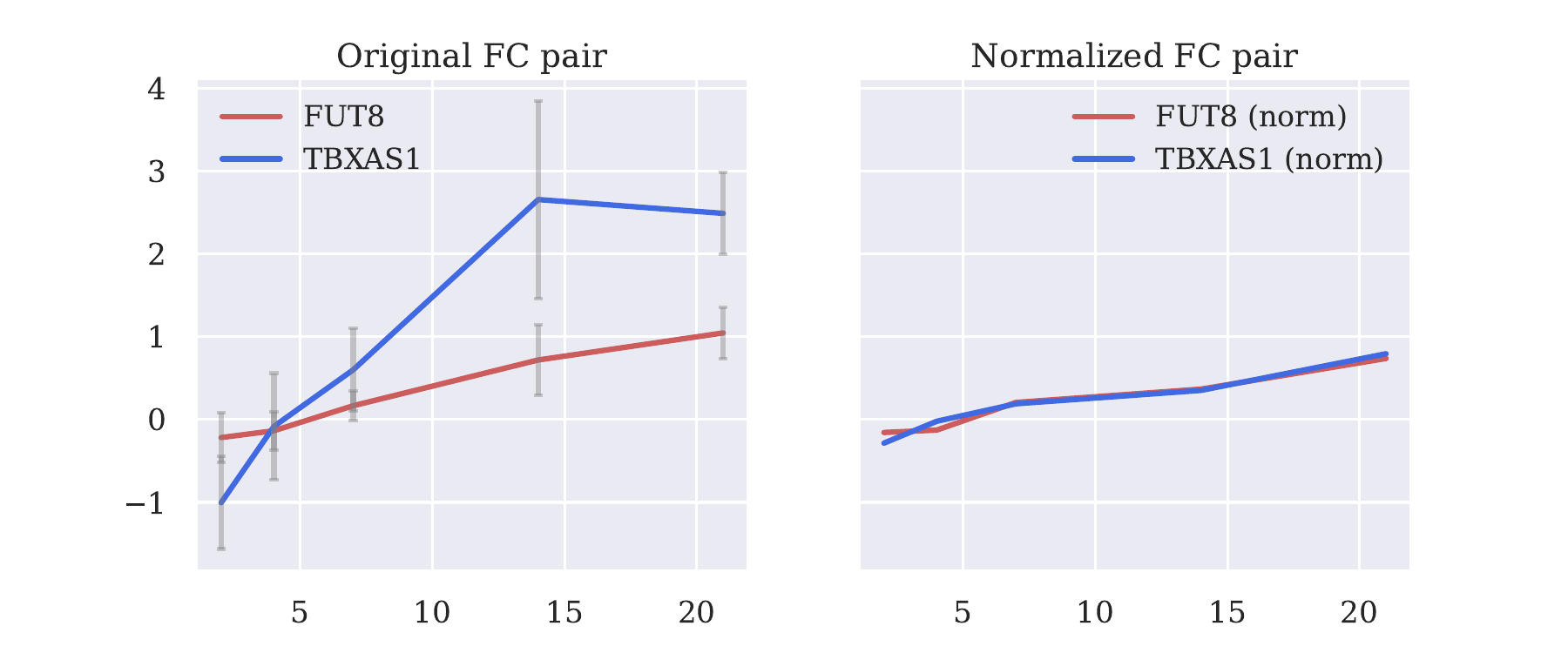}
	\caption{\label{fig:scaling_example}The effect of raw data scaling illustrated on a figure, where means with standard deviation of a pair of transcriptomic fold changes are plotted, inferred from the original data on the left and from scaled data on the right. As a result of scaling, the fold changes of genes TBXAS1 and FUT8 are rendered significantly closer than the original.  For instance,  it can be observed that the original fold changes are both characterized by almost monotonous growth during the whole period after irradiation.  On the one hand, the curve of gene TBXAS1 is more concave, which can be neglected, and the difference is reduced by the scaling with respect to the norm.  On the other hand, the scaling with respect to the standard deviation reduces the peak in the mean of gene TBXAS1 observed at day 14, which is also negligible due to very high standard deviation at that point.}
\end{figure}

The two scaling transformations described above are applied in a consecutive manner: the fold change norm scaling is calculated based on the result of the scaling by standard deviation, which implies that the norm of the final output is equal to 1. Thus, we obtain a processed dataset, from which new pairs of random fold changes estimators are constructed,  and finally the pairwise distances are calculated. An illustrative example for the effect of preprocessing on the fold changes can be found in Figure \ref{fig:scaling_example}.

\subsection{Fold change estimation from preprocessed data}
\label{sec:fc_est}
After applying the preprocessing to the response $Y_{i,k,j,t}$ of an entity $i$ at the time point $t$ for a replicate $j$ under the experimental condition $k$, the response becomes: 
\begin{equation}
	\widetilde{Y}_{i,k,j,t} = \dfrac{Y_{i,k,j,t}}{\sigma_{\Gamma_{i,t}} \times Norm\left(\Sigma_{\Gamma_i}^{-1}\widehat{\Gamma}_i \right)} \text{, where } \sigma_{\Gamma_{i,t}} = \sqrt{\sigma^2_{\Gamma_{i,t}}}.
\end{equation}
We obtain the following expression by applying the norm defined in \eqref{eq:fc_norm} to the fold change $\widehat{\Gamma}_i$ after the scaling by standard deviation:
\begin{equation}
	Norm\left(\Sigma_{\Gamma_i}^{-1}\widehat{\Gamma}_i \right) =  \sqrt{\sum_{l=1}^{p} \left[ \left( \dfrac{\Gamma_{i,t_l}}{\sigma^2_{\Gamma_{i,t_l}}} \right)^2 + 1 \right]}.
\end{equation}

The joint distribution of a fold change pair obtained from the preprocessed data can be rewritten in the following way:
$$
\begin{bmatrix} \widehat{\widetilde{\Gamma}}_i \\  \widehat{\widetilde{\Gamma}}_{i'} \end{bmatrix} \sim \mathcal{N}\left(\begin{bmatrix} \widetilde{\Gamma}_i \\ \widetilde{\Gamma}_{i'}\end{bmatrix} ,\begin{bmatrix} \Sigma_{\widetilde{\Gamma}_i } & \mathrm{P}_{\widetilde{\Gamma}_i \widetilde{\Gamma}_{i'}} \\  \left( \mathrm{P}_{\widetilde{\Gamma}_i \widetilde{\Gamma}_{i'}}\right)^\intercal & \Sigma_{\widetilde{\Gamma}_{i'}} \end{bmatrix}\right) \text{   such that:}
$$

\begin{itemize}
	\item Means for $x\in\{i, i'\}$:
	\begin{equation}
		\begin{split}
			\widetilde{\Gamma}_x  &= \left( \dfrac{\sum_{j=1}^{n_r}(Y_{i,1,j,t_1} - Y_{i,0,j,t_1})}{n_r   \sigma_{\Gamma_{x,t_1}}   Norm\left( \Sigma_{\Gamma_x}^{-1} \widehat{\Gamma}_x \right)} , \dotso ,  \dfrac{\sum_{j=1}^{n_r}(Y_{i,1,j,t_p} - Y_{i,0,j,t_p})}{n_r   \sigma_{\Gamma_{x,t_p}}   Norm\left( \Sigma_{\Gamma_x}^{-1} \widehat{\Gamma}_x \right)}\right) \\&= \dfrac{1}{Norm\left( \Sigma_{\Gamma_x}^{-1} \widehat{\Gamma}_x \right)}   \left(\dfrac{\Gamma_{x,t_1}}{\sigma_{\Gamma_{x,t_1}}}, \dotso , \dfrac{\Gamma_{x,t_p}}{\sigma_{\Gamma_{x,t_p}}}\right),
		\end{split}
	\end{equation}
	
	\item  Covariance matrices for $x\in\{i, i'\}$: $\Sigma_{\widetilde{\Gamma}_x } = \begin{bmatrix} 
		\sigma^2_{\widetilde{\Gamma}_{x,t_1}} & & 0 \\
		& \ddots &  \\
		0 & & \sigma^2_{\widetilde{\Gamma}_{x,t_p}} \end{bmatrix},$
	\begin{equation}\label{eq:fc_norm_cov}
		\begin{split}
			\text{with }\sigma^2_{\widetilde{\Gamma}_{x,t}} = \dfrac{\sum_{j=1}^{n_r}\left[(\widetilde{Y}_{i,1,j,t} - \overline{\widetilde{Y}_{i,1,t}} )^2 + (\widetilde{Y}_{i,0,j,t} - \overline{\widetilde{Y}_{i,0,t}})^2\right]}{n_r - 1} &= \dfrac{\sigma^2_{\Gamma_{x,t}} }{\sigma^2_{\Gamma_{x,t}}   \left(Norm\left( \Sigma_{\Gamma_x}^{-1} \widehat{\Gamma}_x \right)\right)^2} \\ &= \dfrac{1}{ \left(Norm\left( \Sigma_{\Gamma_x}^{-1} \widehat{\Gamma}_x \right)\right)^2} 
		\end{split}.
	\end{equation}
	
	\item  Cross-covariance matrix: $\mathrm{P}_{\widetilde{\Gamma}_i \widetilde{\Gamma}_{i'}} = \begin{bmatrix} 
		\rho_{\widetilde{\Gamma}_{i,t_1}\widetilde{\Gamma}_{i',t_1}}& & 0 \\
		& \ddots & \\
		0 & & \rho_{\widetilde{\Gamma}_{i,t_p}\widetilde{\Gamma}_{i',t_p}}
	\end{bmatrix},$
	\begin{equation}\label{eq:fc_norm_ccov}
		\begin{split}
			\text{with   }	\rho_{\widetilde{\Gamma}_{i,t}\widetilde{\Gamma}_{i',t}}  &= \dfrac{\sum_{j=1}^{n_r}\left[(\widetilde{Y}_{i,1,j,t} - \overline{\widetilde{Y}_{i,1,t}} )(\widetilde{Y}_{i'1j}^t - \overline{\widetilde{Y}_{i',1,t}} ) + (\widetilde{Y}_{i,0,j,t} - \overline{\widetilde{Y}_{i,0,t}} )(\widetilde{Y}_{i'0j}^t - \overline{\widetilde{Y}_{i',0,t}}) \right]}{n_r - 1} \\
			&= \dfrac{\rho_{\Gamma_{i,t}\Gamma_{i',t}} }{\sigma_{\Gamma_{i',t}}   \sigma_{\Gamma_{i',t}}   Norm\left( \Sigma_{\Gamma_{i}}^{-1} \widehat{\Gamma}_{i} \right)   Norm\left( \Sigma_{\Gamma_{i'}}^{-1} \widehat{\Gamma}_{i'} \right)   }.
		\end{split}
	\end{equation}
\end{itemize}

\section{Parameters used to simulate data}\label{sec:sim_params_1}
\subsection{Scenario M1}
\paragraph{Cluster 1:}
$a \sim  \mathcal{N}(0.05, 0.005^2)$, $b \sim  \mathcal{N}(-10a_0, 4a_0^2)$ with $a_0 \stackrel{d}{=} a$, and $c \sim  \mathcal{N}(2, 1)$.

\paragraph{Cluster 2:}
$a \sim  \mathcal{N}(-0.01, 0.001^2)$, $r_1 \sim  \mathcal{N}(5, 1)$, $r_2 \sim  \mathcal{N}(15, 1)$, $c \sim  \mathcal{N}(6a_0, 4a_0^2)$ with $a_0 \stackrel{d}{=} a$, and $d \sim  \mathcal{N}(3, 1)$.

\paragraph{Cluster 3:}
$a \sim  \mathcal{N}(0.01, 0.001^2)$, $r_1 \sim  \mathcal{N}(5, 1)$, $r_2 \sim  \mathcal{N}(15, 1)$, $c \sim  \mathcal{N}(6a_0, 4a_0^2)$ with $a_0  \stackrel{d}{=} a$, and $d \sim  \mathcal{N}(3, 1)$.

\paragraph{Cluster 4:}
$a \sim  \mathcal{N}(5 \times 10^{-3}, (5 \times 10^{-5})^2)$, $r_3 \sim  \mathcal{N}(2, 0.2^2)$, $r_4 \sim  \mathcal{N}(10, 0.5^2)$, $r_5 \sim  \mathcal{N}(18, 0.2^2)$, $b \sim  \mathcal{U}([-0.05, 0.05])$, and $c \sim  \mathcal{N}(2, 0.5^2)$.

\subsection{Scenario M2}
\paragraph{Cluster 1:}
$s \sim  \mathcal{U}([-10, 10])$, $a \sim  \mathcal{N}(0.05, 0.002^2)$, $b \sim  \mathcal{N}(-11a_0, 4a_0^2)$ with $a_0  \stackrel{d}{=} a$, and $c \sim  \mathcal{N}(2, 0.5^2)$.

\paragraph{Cluster 2:}
$s \sim  \mathcal{U}([-10, 10])$, $a \sim  \mathcal{N}(-0.003, (10^{-5})^2)$, $r_1 \sim  \mathcal{N}(8, 1)$, $r_2 \sim  \mathcal{N}(12, 1)$, $c \sim  \mathcal{N}(6a_0, 4a_0^2)$ with $a_0  \stackrel{d}{=} a$, and $d \sim  \mathcal{N}(3, 0.5^2)$.

\paragraph{Cluster 3:}
$s \sim  \mathcal{U}([-10, 10])$, $a \sim  \mathcal{N}(0.003, (10^{-5})^2)$, $r_1 \sim  \mathcal{N}(8, 1)$, $r_2 \sim  \mathcal{N}(12, 1)$, $c \sim  \mathcal{N}(6a_0, 4a_0^2)$ with $a_0  \stackrel{d}{=} a$, and $d \sim  \mathcal{N}(2, 0.5^2)$.

\paragraph{Cluster 4:}
$s \sim  \mathcal{U}([-7, 7])$, $a=\lvert a_0 \rvert$ with $a_0 \sim  \mathcal{N}(2, 1)$, $b \sim  \mathcal{U}([0.3, 0.5])$, and $c \sim  \mathcal{N}(2, 0.5^2)$.

\section{Details on the methods used in simulations}\label{sec:sim_params_2}
\paragraph{Hierarchical clustering (\textit{scikit-learn}):}
type: agglomerative,  linkage: complete.

\paragraph{Dynamic time warping (\textit{dtaidistance}):}
warping window: 3.

\paragraph{Joint clustering and alignment with functional data analysis (\textit{fdacluster}):\\}
centroid\_type = "medoid", warping\_class = "shift", distance\_relative\_tolerance=0.2, warping\_options = c(0.1, 0.1).

\section{Note on Pathway Studio}\label{sec:ps}
Pathway Studio (Web Mammal version 12.4.0.3 )\citep{nikitin_pathway_2003} is a bioinformatic software able to identity biological pathways enriched with selected entities, for example in our case a given gene ou protein list constituting a cluster.
The identification of these pathways is based on an Elsevier knowledgebase of biological relationships called ResNet database compiled by the application of a natural language tool called MedScan on millions of articles and abstracts as well as clinical trials referenced in pubmed.

In Pathway Studio, the pathways statistically enriched in a given list of genes are identified through the Fisher's Exact. The p-values correspond to a right-sided statistical test on a $2\times2$ contingency table to infer if the frequency of genes involved in a given pathway can be considered as significantly associated by more than random chance.

\newpage
\section{Supplementary figures}
\begin{figure}[H]
	\centering
	\includegraphics[clip, trim=0cm 0cm 0cm 1cm,width=0.9\textwidth]{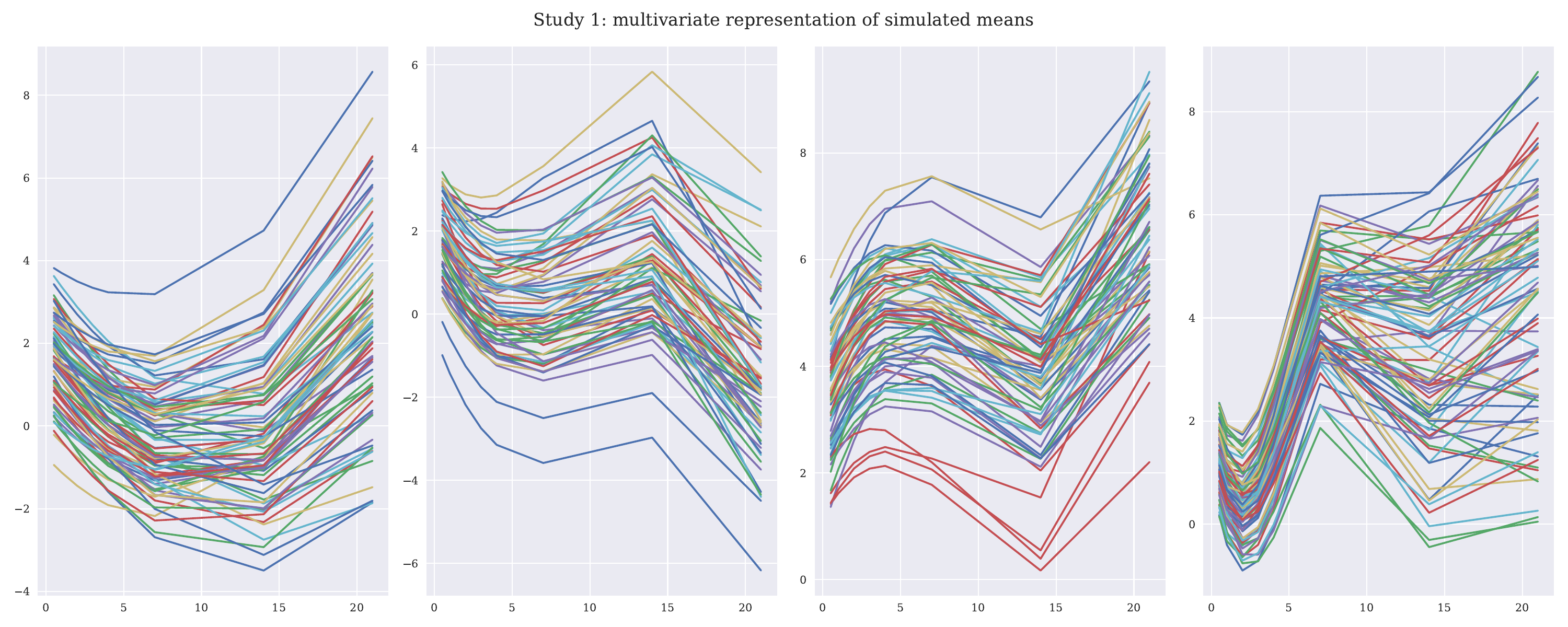}
	\caption{\label{fig:sim1}Multivariate representation of simulated means from each of the four clusters, simulated according to the scenario M1.}
\end{figure}

\begin{figure}[H]
	\centering
	\includegraphics[clip, trim=0cm 0cm 0cm 2cm,width=0.9\textwidth]{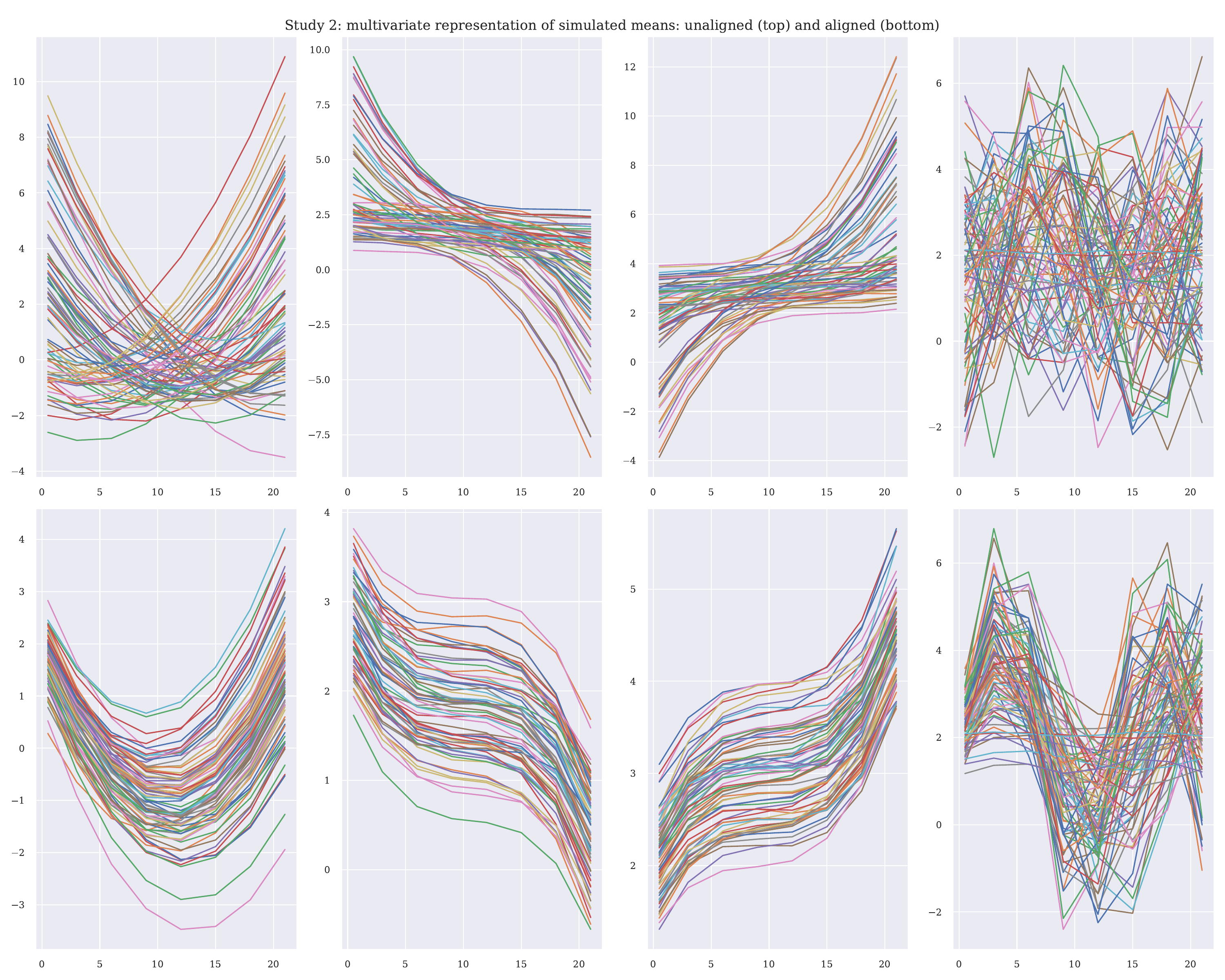}
	\caption{\label{fig:sim2}Multivariate representation of simulated means from each of the four clusters, simulated according to the scenario M2. Top: unaligned means. Bottom: aligned means.}
\end{figure}

\begin{figure}[H]
	\centering
	\includegraphics[clip, trim=2.5cm 1cm 2.5cm 1cm,width=\textwidth]{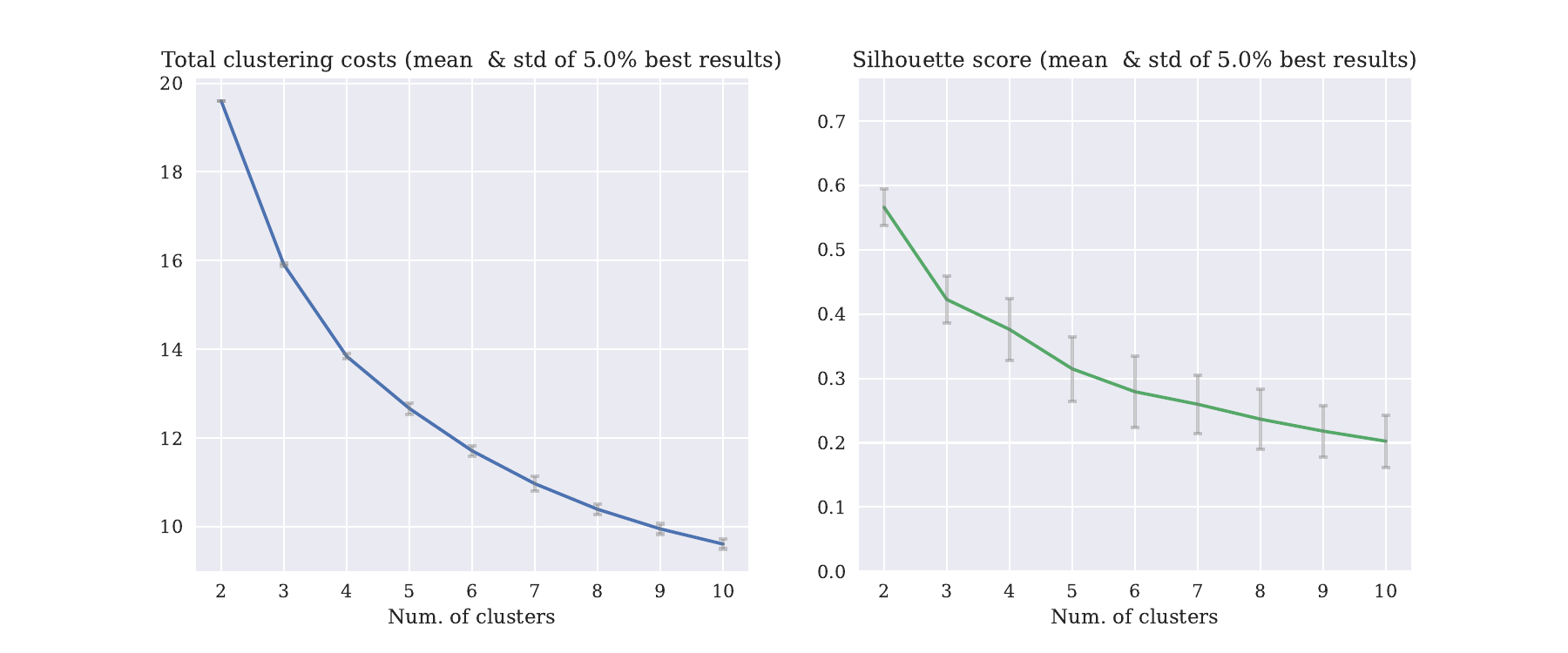}
	\caption{\label{fig:cost}Means of cost and silhouette score with standard deviation of the 20\% best outcomes of clustering of the LINAC dataset with $\mathbf{\widehat{diss}_s}$ k-medoids for the numbers of clusters in the set of fold changes ranging from 2 to 10. The most distinguishable shoulder for the total cost can be observed for 3 clusters, whereas the silhouette score declines for larger number of clusters. Both criteria suggest that smallest number of clusters should be chosen.}
\end{figure}

\begin{figure}
	\centering
	\vspace{-60pt}
	\includegraphics[clip, trim=0cm 3cm 0cm 3cm,scale=0.8]{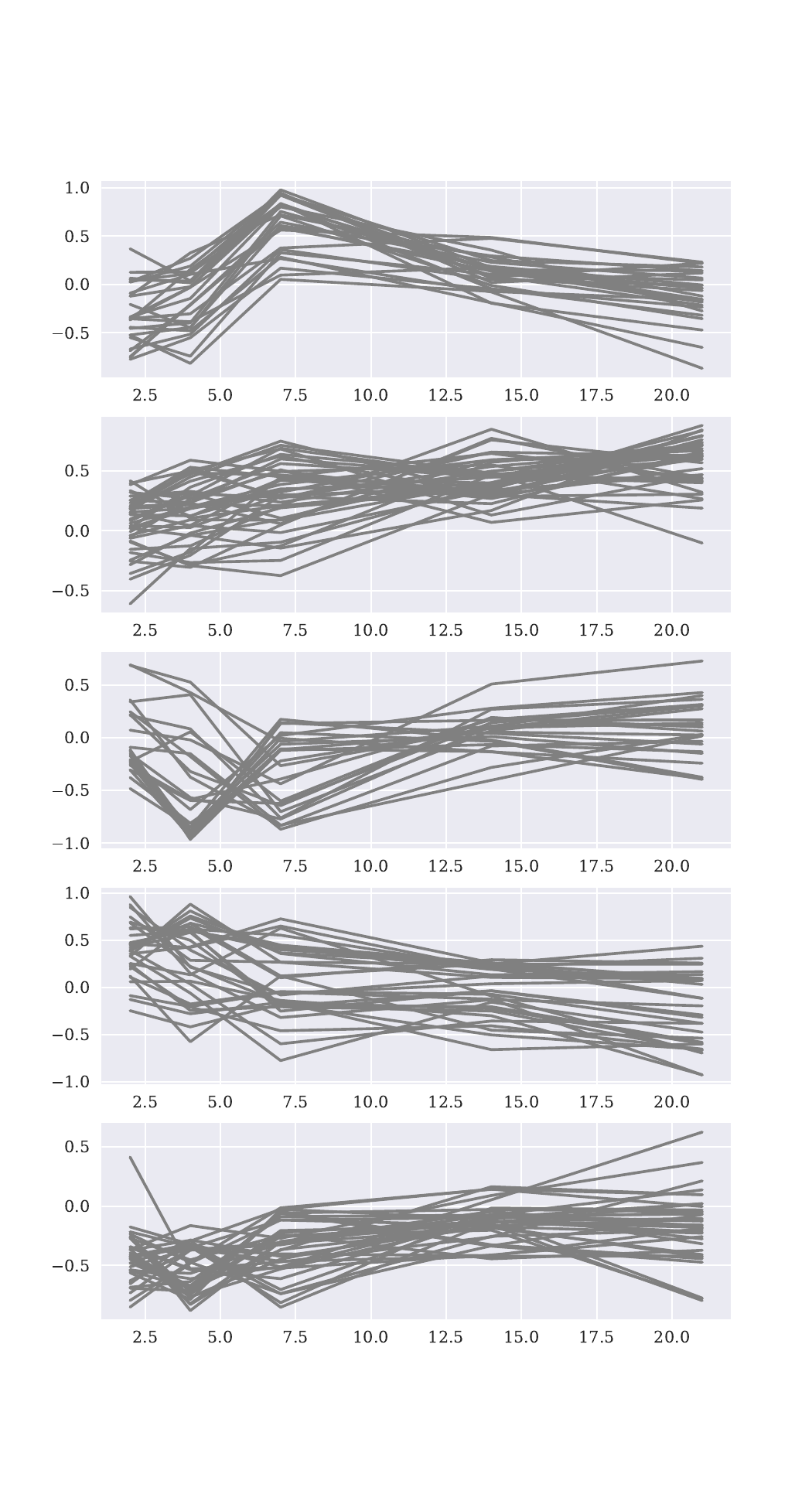}
	\caption{\label{fig:sp_cl}Means of the fold changes in each of the 5 clusters obtained as a result of the clustering of the LINAC dataset with $\mathcal{OWD}$-based spectral clustering. The order of clusters is chosen to maximize the pairwise intersections with the clusters presented in Figure \ref{fig:kmed_alphee_5}.}
\end{figure} 	

\begin{figure}
	\centering
	\vspace{-45pt}
	\includegraphics[width=0.83\textwidth]{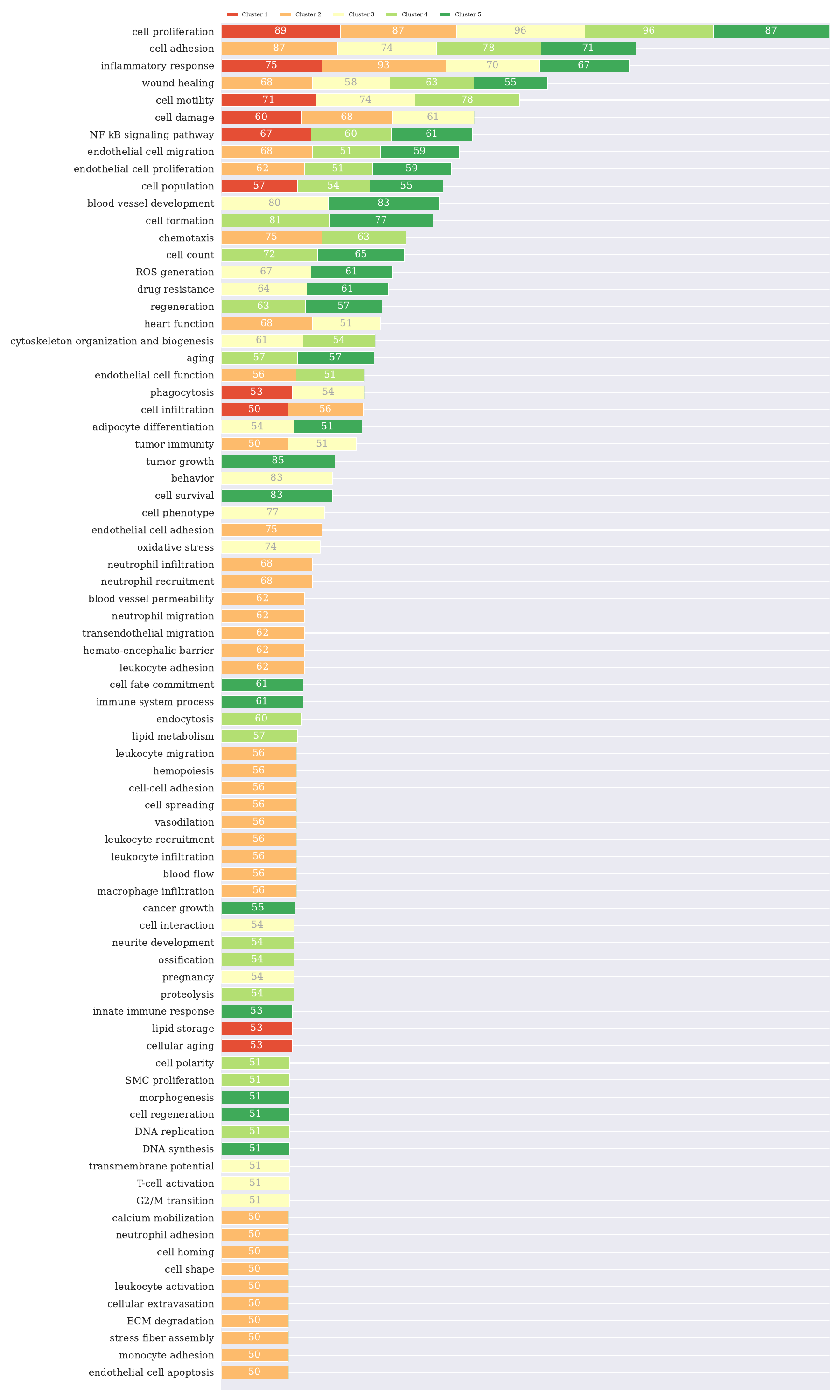}
	\caption{\label{fig:EA_PS}Summary of the enrichment with cellular processes of the clusters obtained for the LINAC dataset. Cellular processes, listed on the left, are sorted from the most represented to the least. The enriched clusters are indicated with different colors. The numbers on the bars indicate the percentage of overlap with the given process for the given cluster.}
\end{figure}

\bibliographystyle{abbrvnatnourl}
\addcontentsline{toc}{chapter}{Bibliography}
\bibliography{paper_1_stats_refs}

\end{document}